\documentclass[aps,physrev,twocolumn,groupedaddress, showkeys, nofootinbib]{revtex4-2}
\usepackage{graphicx}
\usepackage{dcolumn}
\usepackage{bm}
\usepackage{amsmath}
\DeclareMathOperator{\Tr}{Tr}
\usepackage{amssymb} 
\usepackage[table]{xcolor}
\usepackage{float}
\usepackage{graphicx}
\usepackage{tikz}
\usepackage{tabularx}
\usepackage[version=4]{mhchem}
\usetikzlibrary{quantikz} 
\usepackage{array}
\begin{document}


\title{Logical Error Rates for a [[4,2,2]]-encoded Variational Quantum Eigensolver Ansatz}
\thanks{This manuscript has been authored in part by UT-Battelle, LLC under contract DE-AC05-00OR22725 with the U.S. Department of Energy. The United States Government retains and the publisher, by accepting the article for publication, acknowledges that the United States Government retains a non-exclusive, paid-up, irrevocable, world-wide license to publish or reproduce the published form of this manuscript, or allow others to do so, for United States Government purposes. The Department of Energy will provide public access to these results of federally sponsored research in accordance with the DOE Public Access Plan (http://energy.gov/downloads/doe-public-access-plan).}


\author{Meenambika Gowrishankar}
\email[]{Contact Author: mgowrish@vols.utk.edu}
\affiliation{Bredesen Center for Interdisciplinary Research and Graduate Education, University of Tennessee, Knoxville, TN, USA}
\affiliation{Quantum Science Center, Oak Ridge National Laboratory, Oak Ridge, TN, USA}
\author{Daniel Claudino}%
\affiliation{
 Quantum Information Science Section, Oak Ridge National Laboratory, Oak Ridge, TN, USA
}
\affiliation{Quantum Science Center, Oak Ridge National Laboratory, Oak Ridge, TN, USA}

\author{Jerimiah Wright}
\affiliation{
 Quantum Information Science Section, Oak Ridge National Laboratory, Oak Ridge, TN, USA
}
\author{Travis Humble}
\affiliation{Bredesen Center for Interdisciplinary Research and Graduate Education, University of Tennessee, Knoxville, TN, USA}
\affiliation{
 Quantum Information Science Section, Oak Ridge National Laboratory, Oak Ridge, TN, USA
}
\affiliation{Quantum Science Center, Oak Ridge National Laboratory, Oak Ridge, TN, USA}


\date{\today}

\begin{abstract}
Quantum computing offers a potential for algorithmic speedups for applications, such as large-scale simulations in chemistry and physics. However, these speedups must yield results that are sufficiently accurate to predict realistic outcomes of experiments precisely. Delivering on the promise of high accuracy and precision requires methods to evaluate the computational accuracy of the quantum computing devices. We develop a framework to estimate the computational accuracy of near-term noisy, intermediate scale quantum (NISQ) computing devices using a quantum chemistry application. Application benchmarks that run on NISQ devices require techniques for mitigating errors to improve accuracy and precision. We use device agnostic error-mitigation schemes, quantum error detection and readout error detection, with post-selection to mitigate the dominant sources of noise. We evaluate the framework by simulating the ground state of molecular hydrogen with the variational quantum eigensolver (VQE) algorithm, estimating the energy and calculating the precision of the estimate  using numerical simulations with realistic noise models. 
We first quantify the improvement in the logical error rate and state fidelity of the VQE application when encoded with the [[4,2,2]] quantum error detection code. When additionally encoded with readout error detection, we show that compared to the unencoded simulation, the encoded simulation yields a more accurate estimate by more than 1 mHa (0.027 eV) with comparable precision and higher state fidelity. Additionally, unlike the best estimate from the unencoded simulations, the results from the encoded simulation fall within the chemical accuracy threshold of 1.6 mHa of the exact energy. The estimated accuracy and precision indicate that current quantum computers can achieve error rates that yield useful outcomes for chemical applications.
\end{abstract}


\maketitle
\section{Introduction}
Quantum computing hardware has made remarkable strides in improving the coherence times of qubits and the fidelity of gates with error rates in state-of-the-art devices around $0.5-0.01\%$ per two-qubit gate \cite{errorsuppressing, nuetralatom, yamamoto2024demonstrating}. Error rates required for implementing large scale quantum computations are much lower at $\sim$$10^{-9}$, which is considered extremely challenging to achieve on hardware \cite{shorftqc}. Lower error rates are made possible by the application of quantum error correction (QEC), which involve redundantly encoding logical qubits using a larger number of physical qubits such that errors can be detected, decoded and corrected during the computation \cite{shor1995scheme,steane1996error,calderbank1996good,devitt2013quantum}. The leading QEC codes that promise such error rates at physical gate error of $0.1\%$ are the surface code and quantum low-density parity-check (LDPC) codes \cite{fowler, bravyi2024highthreshold}. However, successfully implementing these error correction codes on near-term devices for fault-tolerant quantum algorithms remains an ongoing area of research \cite{errorsuppressing, livingston2022experimental,sivak2023real, ryan2021realization,krinner2022realizing,google2023suppressing, mayer_benchmarking_2024}. 

Quantum error detection (QED) codes, on the other hand, can be implemented on current quantum computers. Unlike QEC, QED uses encoded data qubits only to identify if an error occurred and does not require syndrome decoding or feed-forward operations for its implementation. For example, the [[4,2,2]] QED code uses 4 physical qubits to encode 2 logical qubits and detects at most one physical qubit error \cite{devitt2013quantum}. While QED does not enable fault tolerant operations, post-selection of the flagged measurement results can be used to improve the accuracy of calculated outcomes.  

QED codes have been studied extensively both theoretically and experimentally \cite{leung_approximate_1997,knill2004fault,knill_quantum_2005, reichardt_error-detection-based_2009,zhong_reducing_2014,corcoles2015demonstration,gottesman2016quantum,vuillot2017error,linke2017fault,willsch2018testing,takita_experimental_2017,rosenblum_fault-tolerant_2018,roffe2018protecting,chen_calibrated_2022}.
Most experiments to date have focused on demonstrating that QED codes can successfully detect 
errors during state preparation \cite{corcoles2015demonstration,linke2017fault,roffe2018protecting,zhong_reducing_2014}, while others have tested early fault tolerance on near-term devices \cite{willsch2018testing,vuillot2017error,takita_experimental_2017,rosenblum_fault-tolerant_2018}.
The [[4,2,2]] code and its variant [[4,1,2]] code have been at the center of most of these studies. Several studies have reported on improvements during state preparation, while others have demonstrated the use for applications \cite{errordetection,pokharel2022better,yamamoto2024demonstrating}. These encoded applications include using the [[4,2,2]] QED code with the variational quantum eigensolver (VQE) algorithm and Grover's search algorithm, respectively \cite{errordetection,pokharel2022better}. The former demonstrated the use of the [[4,2,2]] code to encode a two-qubit ansatz for molecular hydrogen for the VQE algorithm \cite{errordetection}. The study showed improvement in accuracy of the ground state expectation value of the encoded hydrogen ansatz across the potential energy surface when compared to the unencoded simulation on an IBM 5Q quantum computing device. The encoding employs a circuit construction introduced in \cite{gottesman2016quantum} that uses an ancilla and a destructive parity measurement to construct state preparation circuits for certain specific input states using the [[4,2,2]] error detection code such that the qubits are fault tolerantly protected. 

In this work, we build upon the results in \cite{errordetection} by developing a framework using classical error detection in addition to QED for effectively managing errors on near-term quantum computers and demonstrate results within the domain specific accuracy threshold. We use `chemical accuracy' as a useful benchmark to evaluate the computational accuracy and error estimates of a near-term quantum computer in applications of computational chemistry. Chemical accuracy is defined in practice to be within $1.6$ mHa of the full configuration interaction (FCI) energy for the chosen one-particle basis set and is necessary for making reliable predictions about the properties derived from these estimates, such as atomization energies and reaction enthalpies \cite{cao_quantum_2019}. Many studies have reported average energy estimates close to or within chemical accuracy for molecular hydrogen using VQE \cite{colless_robust_2018,H2Hamiltonian,omalley_scalable_2016}, however, the precision of these estimates is unclear. We estimate the error rate at which both accuracy and precision of the energy are within chemical accuracy when using this encoding in comparison to the unencoded simulation of the ground state of molecular hydrogen. A recent study reported energy estimates with both accuracy and precision within chemical accuracy for ground state of molecular hydrogen for both the unencoded and the [[4,2,2]]-encoded ansatz on a Quantinuum device \cite{van2024end}. Among the key differences between their framework and ours is their use of a different ansatz, use of additional methods such as classical shadows and simulation of twice the number of samples of the [[4,2,2]]-encoded ansatz as the unencoded ansatz for calculating the energy estimates. 

\section{Background}
This section presents an overview of the VQE algorithm for estimating the energy of a molecular Hamiltonian with chemical accuracy. We then review how to apply the encoding methods defined by the [[4,2,2]] code and the repetition code for creating the variational ansatz state before discussing methods to post-process the detection outcomes. 
\subsection{Molecular Electronic Hamiltonian}
The central problem of electronic structure theory is to find the ground state of a molecule by finding the lowest eigenvalue of the electronic Hamiltonian. We start with the electronic Hamiltonian with coefficients $h_{pq}$ and $h_{pqrs}$ that are a function of the internuclear geometry $R$ and is represented in the second quantized form as: 
\begin{equation}
    H(R) = \sum_{pq} h_{pq} a^{\dagger}_p a_q + \frac{1}{2}\sum_{pqrs}h_{pqrs} a^{\dagger}_pa^{\dagger}_qa_ra_s
\end{equation}
where $a^{\dagger}_p$ and $a_q$ are creation and annihilation operators and follow the cannonical fermionic anti-commutation relations:
\begin{equation}
    \begin{split}
    \{a^{\dagger}_p,a_q\} &= \delta_{pq}\\  
    \{a_p,a_q\} &= 0
    \end{split}
\end{equation}
%
\par 
In the case study below, we estimate the ground state energy of the hydrogen molecule using VQE. The Hamiltonian in the minimal STO-3G basis can be presented in a two-qubit representation that considers only the spin singlet configuration \cite{whitfield_simulation_2011}. 
Using a series of fermionic and spin transformations we arrive at the final Hamiltonian shown in Equation \ref{eq:Ham2}.
\begin{equation}
    \begin{split}
        H(R) &= g_0 I + g_1 Z_0 + g_2 Z_1 + g_3Z_0Z_1 + g_4X_0X_1
    \end{split}
    \label{eq:Ham2}
\end{equation}
Details of the transformations are presented in Appendix \ref{app:hamenc}.
\subsection{Variational Quantum Eigensolver}
The VQE algorithm is a method to estimate the minimal expectation value of a Hermitian operator with respect to a variable pure  quantum state \cite{peruzzo_variational_2014}. It is based on the variational principle from quantum mechanics, which asserts that only the lowest eigenstate, aka ground state, of a non-negative, H-9ermitian 
operator minimizes the expectation value. Estimating the energy is then described by the optimization 
\begin{equation}
    E(\theta^{\star}) = \min_{\theta}{ \langle \psi(\theta)|H|\psi(\theta)\rangle}
\end{equation}
where $|\psi(\theta) \rangle = U(\theta) | \psi(0) \rangle $ is a variable pure quantum state prepared by a unitary ansatz operator $U(\theta)$ from the reference state $|\psi(0)\rangle$. The parameter $\theta^{\star}$ denotes the optimal value obtained from minimizing the energy.
The VQE reference state is typically represented by the Hartree-Fock solution to the electronic Hamiltonian. In our encoding of the hydrogen molecule above, the Hartree-Fock state corresponds to the $|00\rangle$ state. 
\par
\begin{figure}
    \centering
        \includegraphics[width =0.8\linewidth]{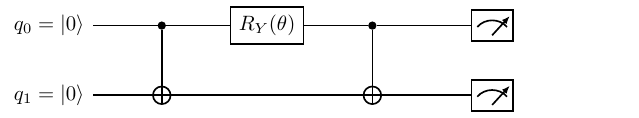}
    \caption{The unencoded two-qubit UCC single parameter VQE ansatz.}
    \label{fig:unenccirc}
\end{figure}

We consider an ansatz based on unitary coupled cluster (UCC) theory which derives from the well known coupled cluster method in quantum chemistry \cite{romero_strategies_2018, bartlett1989alternative,taube2006new}.
The unitary operator approximated to the first Trotter step ($t=1$) yields the UCC singles and doubles ansatz (UCCSD).
%
In the case of molecular hydrogen, the singles excitation operator does not contribute to the energy due to Brillouin's theorem \cite{szabo_modern_1996}. Considering only the contribution from the double excitation $T_2$, shown in Equation \ref{eq:exop} in the appendix, we arrive at the UCC doubles (UCCD) ansatz shown in Equation \ref{eq:unitary}. The transformation leading to this operator is presented in Appendix \ref{app:unitop}.
\begin{equation}
U(\theta) = e^{-i\theta Y_0X_1}
\label{eq:unitary}
\end{equation}
%
The circuit decomposition of Equation \ref{eq:unitary} involves the subcircuit for the $e^{-i\theta Z_0Z_1}$ operator surrounded by single qubit rotations to perform the transformation to $e^{-i\theta Y_0X_1}$\cite{Nielsen_Chuang_2010,barenco1995elementary}. In our encoding, some of the single qubit rotations in the decomposition are redundant for the initial Hartree-Fock input state, $|00\rangle$ and are removed. This results in the two-qubit unencoded circuit shown in Figure \ref{fig:unenccirc}, which prepares the final state $|\chi(\theta)\rangle$ shown in Equation \ref{eq:2qfinstate}.  
\begin{equation}
    |\chi(\theta)\rangle = \cos\frac{\theta}{2}|00\rangle + \sin\frac{\theta}{2}|11\rangle
    \label{eq:2qfinstate}
\end{equation}
\par 
We present results for the expectation value of the Hamiltonian at the internuclear distance of 0.74Å shown in Equation \ref{eq:hamexact}, where the coefficients $g_i$ are obtained from \cite{H2Hamiltonian}. 
\begin{equation}
    \begin{split}
        H &= -0.349833 \ \mathbb{I} - 0.388748 \ Z_0 - 0.388748 \ Z_1 \\& + 0.0111772 \ Z_0Z_1  + 0.181771 \ X_0X_1
    \end{split}
    \label{eq:hamexact}
\end{equation}
The analytical energy expectation value and optimal parameter for this Hamiltonian and ansatz are $E(\theta^{\star})=-1.13712 \ \text{Ha}$ and $\theta^{\star}=-0.22967 \ \text{rads}$, respectively. 

\subsection{[[4,2,2]] Quantum Error Detection Code}   
\par
\begin{figure}
    \centering
    \includegraphics[width = 0.9\linewidth]{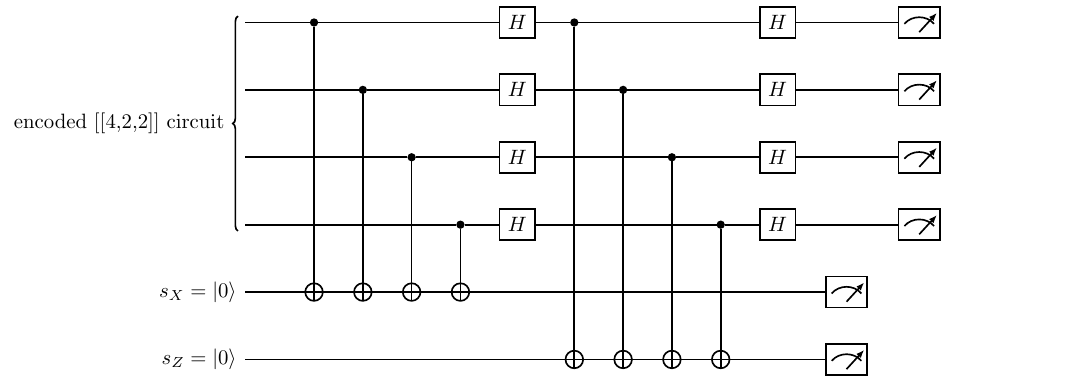}
    \caption{Syndrome measurement circuit for the [[4,2,2]] error detection code.}
    \label{fig:stab}
\end{figure}

The [[4,2,2]] QED code, following the $[[n,k,d]]$ convention, is a distance $d=2$ code that encodes $k=2$ logical qubits using $n=4$ physical qubits. The basis states and operations for the code are provided in the Appendix \ref{app:basisst}. The code 
can detect at most one-single physical qubit bit-flip and/or phase-flip error (Pauli X or Z error, respectively) that occurs during the encoding of the initial state \cite{devitt2013quantum}. The circuit for the error detection or syndrome measurement for this encoding is presented in Figure \ref{fig:stab}. Ancillas $s_X$ and $s_Z$ are used for error syndrome measurements and detect bit-flip and phase-flip errors, respectively. 
\par 
All the basis states of this encoding have even parity (even number of 1s in the physical basis), and any single physical qubit bit-flip error will take the state outside the logical codespace and result in a state with odd parity.
As a result, a single physical qubit bit-flip error will lead to a measurement of $s_X=1$. If the state is rotated using the Hadamard ($H^{\otimes 4}$) operation before measurement, any single physical qubit phase-flip error will lead to a measurement of $s_Z=1$.  Two physical qubit bit-flip errors take the state to another logical state and will remain undetected by the ancillas. Two physical qubit phase-flip errors can add a global phase to the state or leave the state effectively unchanged.
\par
An alternative method to observe the detectable errors is by measuring all the data qubits and selecting joint measurements based on parity. Odd parity measurements indicate that a single qubit bit-flip error has occurred and is akin to performing the stabilizer measurement with ancilla $s_X$ in Figure \ref{fig:stab}. Similarly, odd parity of measurements made after rotating the state by applying the Hadamard ($H^{\otimes 4}$) operation indicate that a physical qubit phase-flip error has occurred and is equivalent to an $s_Z=1$ measurement in Figure \ref{fig:stab}.
The advantage of this method over the stabilizer measurement is the reduction in the number of two-qubit gates for syndrome detection. The disadvantage is that the detection of errors requires a destructive measurement and only allows detection of one of the two types of errors, a physical bit-flip or phase-flip error. 
\par
This method can be coupled with an ancilla for additional error detection during state preparation. This is applicable only to certain specific input logical states, $|\overline{00}\rangle, |\overline{0+}\rangle, |\overline{00}\rangle+|\overline{11}\rangle$ and can be used during the preparation of the Hartree-Fock initial state, $|\overline{00}\rangle$, in this study\cite{gottesman2016quantum}. 
The ancilla detects physical qubit bit-flip errors during preparation of the input state while the parity check measurement described earlier enables detection of single physical qubit bit- or phase-flip errors. 
\subsection{Readout Error Detection (RED)}
\par
The leading single source of error in all quantum computing technologies is readout error, also called measurement error. Since the read-out error needs to be mitigated before applying the post-selection strategies associated with the [[4,2,2]]-encoding, we require a readout error mitigation technique that can be applied to every shot or circuit measurement. Considering that read-out errors are comprised of only bit-flip errors, they can be managed using classical error detection codes. The classical repetition code is a method both for detecting classical bit-flip errors and to filter noisy results from every measurement prior to post-selection for the [[4,2,2]]-encoded ansatz. 
\par
We consider the three-qubit repetition code referred to as [3,1]-RED code, which encodes one logical readout qubit with three physical qubits \cite{hicks_active_2022}. The circuit encodes the data qubits just prior to their measurement. The data qubits are encoded by entangling each of them with two ancillas and errors are detected based on unanimous vote.
\par
The success of the code in mitigating readout errors depends on the difference in magnitude of error rates between the two-qubit gate error and readout error. Having a much lower two-qubit gate error rate than readout error rate will lead to higher accuracy of the observable calculated from the post-selected measurements. A much lower two-qubit gate error rate would introduce much fewer errors due to the gate than the readout error, increasing the probability that the error observed during RED is due to measurement error rather than two-qubit gate error. Using the RED method with a higher two-qubit gate error rate than read-out error rate will lead to a decrease in accuracy of the results as the likelihood that the errors in measurement are caused due to the additional two-qubit gates used for the RED encoding becomes higher than errors introduced due to the measurement itself.  
\section{Encoding Methods}
In this section, we present details on the methods to encode the VQE algorithm and UCCD ansatz using the [[4,2,2]] code. We also review the methods used to simulate and post-process results for the electronic Hamiltonian of molecular hydrogen in the presence of circuit noise.
\subsection{[[4,2,2]] Encoding of the UCC Ansatz}
\begin{figure}
    \centering
        \includegraphics[width =1\linewidth]{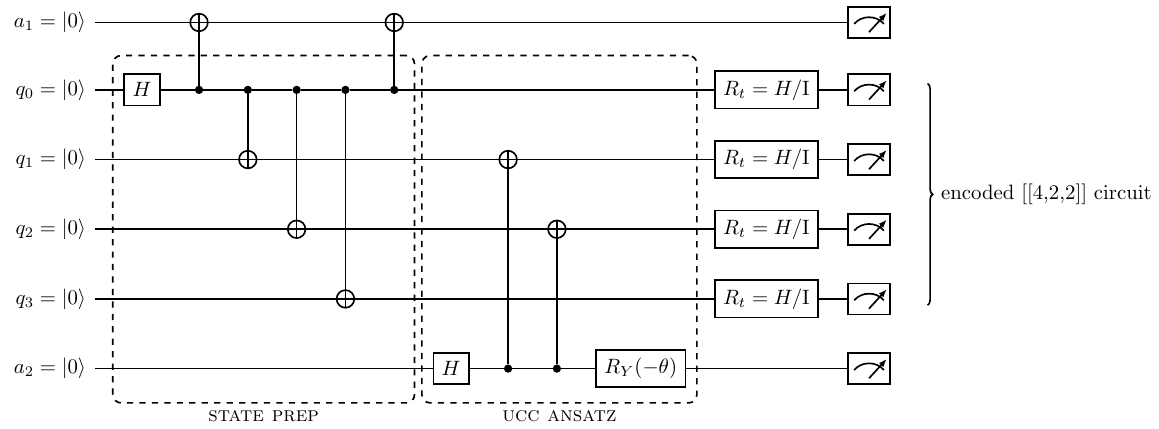}
        
    
    \caption{The [[4,2,2]] encoded ansatz for the hydrogen molecule. The block labeled 'STATE PREP' describes the circuit to prepare the initial state in the [[4,2,2]] encoding. The block labeled 'UCC ansatz' describes the implementation of the UCC derived hydrogen VQE ansatz in the [[4,2,2]] encoding. $R_t = H/I$ indicates the basis of the state prior to measurement, in the $X$ or computational basis, respectively.}
    \label{fig:enccirc}
\end{figure}
\par
The corresponding [[4,2,2]] encoding of the parameterized UCC two-qubit ansatz shown in Figure \ref{fig:unenccirc} for simulating the hydrogen molecule is presented in Figure \ref{fig:enccirc}
\cite{errordetection}. The first block on the left, labeled ``STATE PREP", represents the circuit to prepare the initial logical $|00\rangle$ state of the unencoded ansatz and can be prepared in a wide variety of ways \cite{gottesman2016quantum,errordetection,vuillot2017error}. The second block labeled,``UCC ANSATZ", represents the execution of the UCC ansatz in the [[4,2,2]] encoding \cite{errordetection}. The parameterized rotation gate in the ansatz is non-transversal in this encoding and therefore, an ancilla, $a_2$, is used to teleport the gate. 
The final state following the action of the paramatrized unitary operator corresponding to the encoded circuit ($|\psi_{enc}(\theta)\rangle$) is shown in Equation \ref{eq:finstate}. 
\begin{equation}
    \begin{split}
        |\psi_{enc}(\theta)\rangle &= \frac{1}{\sqrt{2}} [(|0\rangle_{a_1} \otimes(\cos\frac{\theta}{2} |\overline{00}\rangle_{q_0-q_3} \\&+ \sin\frac{\theta}{2} |\overline{11}\rangle_{q_0-q_3} )\otimes |0\rangle_{a_2})\\&+(|0\rangle_{a_1}\otimes(\cos\frac{\theta}{2} |\overline{11}\rangle_{q_0-q_3} \\&- \sin\frac{\theta}{2}|\overline{00}\rangle_{q_0-q_3} )\otimes |1\rangle_{a_2})]
        \end{split}
        \label{eq:finstate}
\end{equation}
\par
The action of this non-transversal gate in the encoded circuit results in a uniform superposition of equivalent states, each representing the unencoded final state, with equal probability to be in $a_2 = |0\rangle$ or $|1\rangle$, and with each minimizing the expectation value at a different parameter. The optimal parameter, $\theta^*$, at $a_2 = |0\rangle$, is shifted by $\pi$ when $a_2=|1\rangle$. For the purpose of calculating expectation values, we only use outcomes with $a_2=|0\rangle$ resulting in the final state shown in Equation \ref{eq:a2state}. This excludes $\sim50\%$ of the total number of outcomes measured, belonging to the subset with $a_2=|1\rangle$, from our calculations. The corresponding state for $a_2 = |1\rangle$ is shown in Equation \ref{eq:a21state}.
\begin{eqnarray}
        |\psi_{enc}^{a_2 = 0}(\theta)\rangle =& |0\rangle_{a_1}\otimes (\cos\frac{\theta}{2} |\overline{00}\rangle_{q0-q_3} \nonumber \\ +& \sin\frac{\theta}{2} |\overline{11}\rangle_{q_0-q_3})\otimes |0\rangle_{a_2} 
    \label{eq:a2state}
\end{eqnarray}
\begin{eqnarray}
        |\psi_{enc}^{a_2=1}(\theta)\rangle &= |0\rangle_{a_1}\otimes (\cos\frac{\theta}{2} |\overline{11}\rangle_{q_0-q_3} \nonumber \\ &-\sin\frac{\theta}{2}|\overline{00}\rangle_{q_0-q_3})\otimes|1\rangle_{a_2}
    \label{eq:a21state}
\end{eqnarray}
\par
For expectation value calculations, since each Pauli term in the Hamiltonian in Equation \ref{eq:Ham2} is in a single basis, $X$ or $Z$, we measure the final state of the encoded ansatz in the computational basis and rotate the state using the Hadamard operation prior to measurement for the ``$X_0X_1$" term. This is shown in Figure \ref{fig:enccirc} as $R_t = H$ for the ``$X_0X_1$" Pauli term or $R_t = I$ for terms in the computational basis, prior to measurement. 
\subsection{RED Encoding of the Unencoded and [[4,2,2]]-Encoded UCC Ansatz}
 
\par
\begin{figure}
    \centering
    \includegraphics[width = 0.8\linewidth]{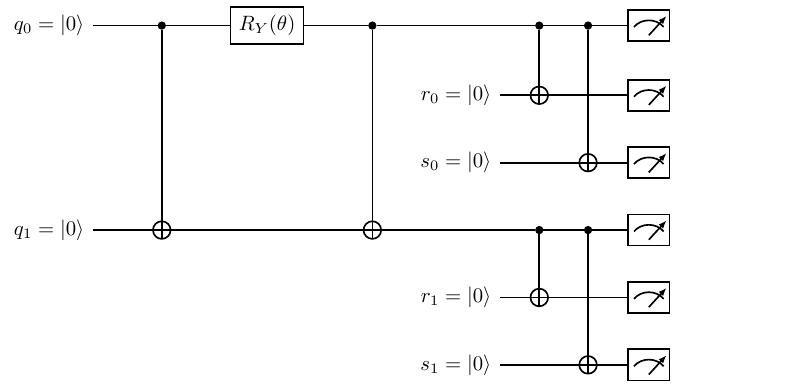}
    \caption{Unencoded UCC ansatz with encoding for [3,1]-RED using qubits $r_i,s_i \ \text{for} \ i \in [0,1]$ for VQE}
    \label{fig:roed_unenc}
\end{figure}
\begin{figure}
    \centering
    \includegraphics[width = 0.98\linewidth]{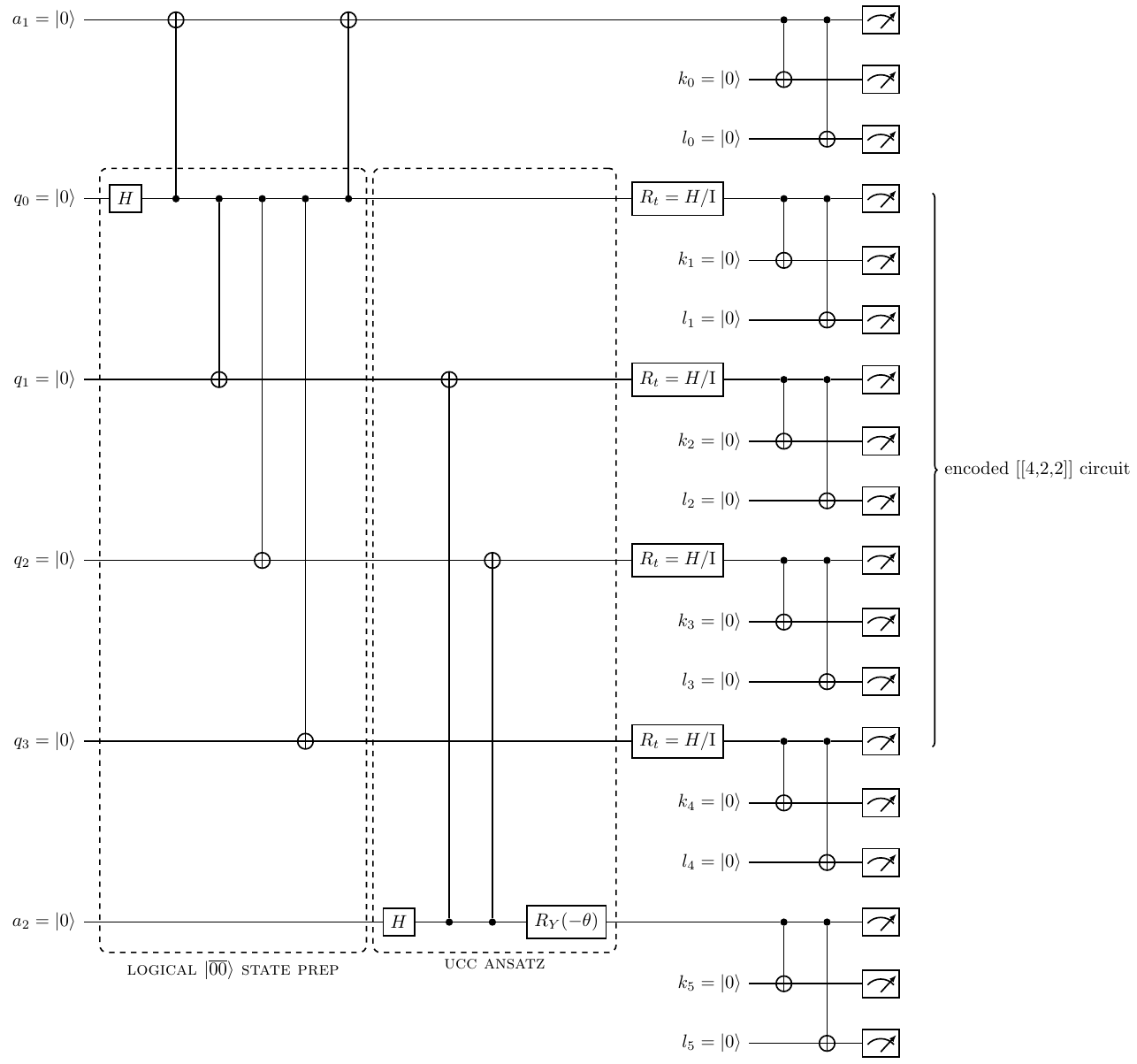}
    \caption{Encoded UCC ansatz using the [[4,2,2]] QED code, and additional encoding with qubits $k_il_i \ \text{for}\ i \in [0,5]$ for [3,1]-RED}
    \label{fig:roed_enc}
\end{figure}

The simulations on the device emulator with the device error model require mitigation of readout error in addition to gate error.  We encode the two circuits, the unencoded two-qubit ansatz and the [[4,2,2]]-encoded ansatz, with error detection code for RED. 
The corresponding circuits with the [3,1]-RED code are shown in Figures \ref{fig:roed_unenc} and \ref{fig:roed_enc}, respectively. The ancillary qubits for readout for the unencoded ansatz are represented by $r_i,s_i$ for data qubits $q_i$ for $i\in[0,1]$. The ancillary qubits for the [[4,2,2]]-encoded ansatz are represented by $k_il_i$ for $i\in [0,5]$. 
\vspace{-5mm}
\section{Error Models and Simulations} \label{numsim}
We use the XACC framework to run the numerical simulations of the VQE algorithm. XACC is a software framework for the development of hardware-agnostic programs for quantum-classical hybrid algorithms, and their implementation on near-term quantum hardware \cite{mccaskey2020xacc,backend-agnostic_2022} . We use the IBM Aer simulator to evaluate the influence of depolarizing noise on the energy accuracy, and Quantinuum H1-1E emulator to evaluate more realistic effects of hardware noise on fully corrected calculations, within XACC, by numerically simulating each ansatz. 
\subsection{Depolarizing Error Model}
\par 
For evaluating the impact of the [[4,2,2]]-encoded ansatz in the presence of gate noise, we run numerical simulations under a standard depolarizing error channel given as
    \begin{equation}
        \xi^{G}_{j}(\rho) = (1-p)\rho' + p \sum_{k} \sigma^{k}_{j} \rho' \sigma^{k}_{j}
    \end{equation}
where $\rho' = G \rho G^{\dagger}$ and $G$ is any single-qubit gate in the circuit acting on qubit $j$ in density matrix $\rho$, $\sigma_{j} \in \{X_j, Y_j, Z_j\}$, and $p$ is the noise parameter. Errors on a two-qubit gate acting on qubits $i$ and $j$ are modeled as $\xi^{G}_{i,j}(\rho) = \xi^{I}_{i}(\xi^{G}_{j}(\rho))$ with $I$ the identity. The noise parameter for the two-qubit gate is an order of magnitude higher than the single-qubit noise parameter, as is typical in current devices. 
The errors resulting from this channel are single qubit errors, $E_{j}$ on qubit $j$ or two qubit errors $E_i\otimes E_j$ on qubits, $i$ and $j$ for $E_k \in \{X_k, Y_k, Z_k, \mathbb{I}_k\}$. 
\par 
We calculate the energy expectation values for both the unencoded and encoded simulations and for each post-selection method, using $N=2\times10^5$ shots, which we determined by calculating the shot count required to estimate the energy within standard error of the mean (SEM) of 0.5mHa for a noiseless simulation. 
We analytically calculated the variance of the Hamiltonian to be $0.04700\ \text{Ha}^2$ by using the expectation value of each Pauli term in Equation \ref{eq:hamexact}. We found the number of shots required to reach our target SEM of $0.5 \ \text{mHa}$ to be $188000$ and rounded up to result in $N =2\times10^5$ shots. We find the minimum energy by scanning 150 values of $\theta \in [-\pi,\pi ]$. 
\subsubsection{\label{subsubsection:PSStrat}Post-Selection Strategies}
\par
We introduce and analyze several different post-selection strategies. We post-select outcomes from the measurement bitstrings once all qubits are measured. The expectation values of each Pauli term in the Hamiltonian are estimated from the available, post-selected measurements. Measurement of ancilla, $a_1=1$ indicates a bit-flip error has occurred on qubit $q_0$. As a result, for post-selection by ancilla $a_1$ (PSA) measurement, we discard all measurements with $a_1=1$.
\par 
Since measurements of the encoded qubits/data qubits with odd parity indicate a single physical qubit bit- or phase-flip error, for post-selection by logical state parity (PSP), we discard measurement bitstrings that have odd parity but include measurement counts with both measurements of ancilla, $a_1=0$ and $a_1=1$. 
We also consider a post-selection strategy labeled PSAP that combines both the PSA and PSP strategies. In this post-selection method, we discard measurement outcomes that have odd parity or ancilla, $a_1=1$. 
\par 
We also calculate the SEM for the energy estimated from each post-selection method as an estimate of the precision of the calculation. 
\subsubsection{Probability of Success} We report on the probability of success (POS) defined as the fraction of samples $\eta$ that is retained after post-selection for each post-selection strategy.
%
 %
 The SEM for the resulting binomial distribution for each post-selection method is calculated as
\begin{equation}
    \sigma_{\text{POS}} = \sqrt{\frac{(1-\eta)(\eta)}{N}}
    \label{eq:ps_sem}
\end{equation}
where $N$ is the number of samples with $a_2 = 0$. 

\subsubsection{Logical Fidelity and Logical Error}
\par
We calculate the logical fidelity of the prepared states as generated by the noisy circuit simulations. 
We simulate the two-qubit UCCD ansatz state for the hydrogen molecule and the corresponding [[4,2,2]] encoded circuit described in Figures \ref{fig:unenccirc} and \ref{fig:enccirc}, respectively.
With the resulting final state for the encoded circuit shown in Equation \ref{eq:a2state} we use projection operators to assess the impact of each post-selection technique using fidelity. We calculate the fidelity for the PSA method by using operators $\Pi_A$ to project states from the noisy state with $a_1 = |0\rangle$ and identity ($\mathbb{I}$) for all other qubits. For the PSP strategy, we project states within the codespace, i.e., within the states in Equation \ref{eq:basisstate}, using operators $\Pi_{P}$. And we use the operator, $\Pi_{AP}$ for projecting states with $a_1=|0\rangle$ and states within the codespace. The operators are presented in Appendix \ref{app:projexp}. 


\par
We calculate the logical fidelity of the states projected using operators $\Pi_A, \Pi_{P}, \Pi_{AP}$, and the original encoded and unencoded states. We define the fidelity $F$ between two states $\rho_1$ and $\rho_2$ as
\begin{equation}
    F(\rho_1, \rho_2) = \left(\Tr\sqrt{\sqrt{\rho_2} \rho_1 \sqrt{\rho_2}}\right)^2
    \label{eq:fidelity}
\end{equation}
%
Here, we consider the case that $\rho_1 = |\Psi\rangle\!\langle\Psi|$, where $|\Psi\rangle$ is the expected ground state, i.e., the noiseless state, from the noise-free simulation and $\rho_2$ is the representation of the prepared, noisy unencoded, encoded or projected state. The fidelity of the projected state is calculated by $F_i(\rho_1, \rho_i)$, 
where $\rho_i = \frac{\Pi_i\rho_{\text{noisy}}\Pi_i^{\dagger}}{\Tr(\Pi_i\rho_{\text{noisy}}\Pi_i^{\dagger})}$ is the projected state,  $\rho_{\text{noisy}}$ is the noisy encoded state, and $\Pi_i \in \{\Pi_A, \Pi_P, \Pi_{AP}\}$.
We also calculate the minimum expectation value of energy for the state $\rho_i$ 
as 
\begin{equation}
    E(\theta^*) = \Tr(H\rho_i(\theta^*))
    \label{eq:dmenergy}
\end{equation}
which represents an estimate in the limit of infinite samples of the measured state.
\par
As part of the error analysis of the [[4,2,2]] encoded circuit, we calculate the probability of logical error. We define logical error ($p_{\epsilon}$) for the encoded circuit in Figure \ref{fig:enccirc} as probability of any error that takes the target or ideal encoded logical state to a different encoded logical state. We restrict the calculation to states with $a_2=0$. Therefore the logical error is the probability of measuring any state within the codespace shown in Equation \ref{eq:basisstate} in the Appendix other than the state $|\psi_{enc}^{a_2=0}(\theta^*)\rangle$ shown in Equation \ref{eq:a2state}. 


We first calculate the probability to measure the ideal state, $p_{\text{ideal}}$, in the noisy mixed state, $\rho_{\text{noisy}}$, and the probability to measure any logical state, i.e., any state in the codespace described in Equation \ref{eq:basisstate}. 
\begin{subequations}
    \begin{align}
        p_{\text{ideal}} &= \Tr(\rho_{\text{noisy}}\rho_{\text{ideal}})\\
        p_{\text{logical}} &= \Tr(\Pi_{P}\rho_{\text{noisy}})\label{eq:plogstate}
    \end{align}
\end{subequations}
From these we find the probability of any error ($p_{\epsilon_{all}}$) and probability of any non-logical error ($p_{\epsilon_{NL}}$), which is any error, single- or multi-qubit, that takes the target or ideal state, $\rho_{\text{ideal}}$, outside the codespace described in Equation \ref{eq:basisstate}.
\begin{subequations}
    \begin{align}
        p_{\epsilon_{all}} &= 1-\Tr(\rho_{\text{noisy}}\rho_{\text{ideal}})\\
        p_{\epsilon_{NL}}&= 1-p_{\text{logical}}
    \end{align}
\end{subequations}
The probability of any logical error to occur, $p_{\epsilon_L}$, is then obtained by subtracting the probability of any non-logical error from the probability of any error as shown in Equation \ref{eq:logerrdef}.
\begin{equation}
    \begin{split}
        p_{\epsilon_{L}} &= p_{\epsilon_{all}}-p_{\epsilon_{NL}}
    \end{split}
    \label{eq:logerrdef}
\end{equation}

We additionally calculate the impact of PSA on the logical error by replacing $\Pi_{P}$ in Equation \ref{eq:plogstate} with $\Pi_{AP}$ and find probability of logical error $p_{\epsilon_A}$ in this case as:

\begin{equation}
    p_{\epsilon_A} = \Tr(\Pi_{AP}\rho_{\text{noisy}}) - \Tr(\rho_{\text{noisy}}\rho_{\text{ideal}})
\end{equation}
\subsubsection{Circuit Error Analysis}
\par

\par
Errors detected by the PSP method are readily interpreted as single physical qubit bit-flip or phase-flip error. However, the errors detected by the $a_1$ in the PSA and PSAP methods require additional analysis. 
Errors under the depolarizing noise model we have considered include one- and two-qubit gate errors of the form $E_i$ on qubit $q_i$ and $E_i \otimes E_j$ on qubits $q_i,q_j$, respectively. Additionally, $CNOT$ gates generate two-qubit correlated errors such as a bit-flip error on the control qubit will result in a bit-flip error in the target qubit. The effective error ($E^{'}_{k}$) after a CNOT operation is applied on control qubit, $i$ and target qubit $j$ after an error has occurred on one of the qubits can be represented as $U_{ij}E_{k}|q_{i}q_{j}\rangle = E^{'}_{k} U_{ij}|q_{i}q_{j}\rangle$ where $U_{ij}=CNOT_{ij}$ is the ideal operation and $E^{'}_{k}=U_{ij}E_{k}U^{\dagger}_{ij}$ for $k\in\{i,j\}$. Multi-qubit errors of the form $E_1\otimes E_2\otimes E_3\otimes E_4$ for $E_k \in \{X, Y, Z\}$ may also arise from concurrent one-qubit noise processes occurring on multiple qubits or from propagation of error under the two-qubit error model. 
\par
\begin{figure}
    \centering
    \includegraphics[width = 0.99\linewidth]{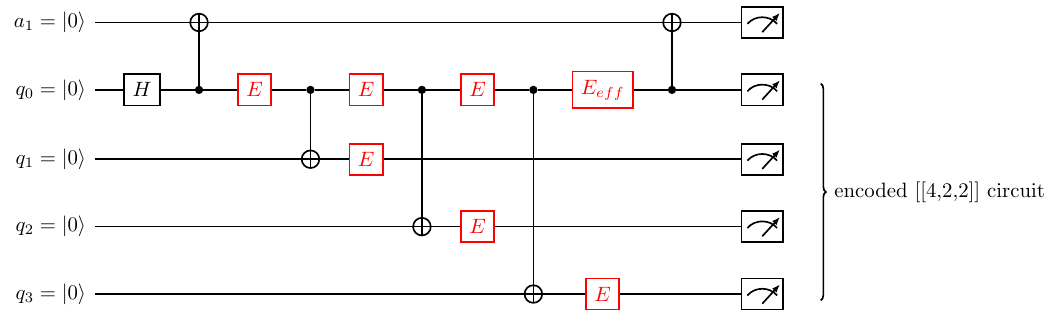}
    \caption{Logical $|\overline{00}\rangle$ state preparation circuit for error analysis. Errors, $E \in \{X,Y,Z\}$, that can occur in the input state preparation circuit, locations are representative. Ancilla $a_1 =1$ if $E_{eff} \in \{X,Y\}$.}
    \label{fig:errorcirc}
\end{figure}

During the [[4,2,2]] input state encoding, as in the section labeled ``STATE PREP" in Figure \ref{fig:enccirc} and shown explicitly in Figure \ref{fig:errorcirc}, the $a_1$ detects any bit-flip error $E_0\in \{X_0,Y_0\}$ that occurs on $q_0$, and effectively remains a bit-flip error, 
$E_{eff} \in \{X_0,Y_0\}$ in Figure \ref{fig:errorcirc}, and results in a measurement outcome for $a_1=1$. 
However, a similarly occurring phase-flip error on $q_0$ will not impact the $a_1$ and therefore, will remain undetected. Since the ``STATE PREP" section is dominated by two-qubit CNOT gates, there's a high likelihood of two-qubit errors occurring in that section. Additionally, these errors may cascade into multi-qubit errors with each successive CNOT gate execution as represented in Figure \ref{fig:errorcirc}. 

\par
Based on the noise processes and resulting errors described, the discarding of the errors by $a_1$, while effected by detection of bit-flip errors on $q_0$, may inadvertently remove multi-qubit errors as well. Construction of the circuit such that all the CNOT gates originate at $q_0$ with $q_0$ as the control ensures that many, if not most, two-qubit errors will impact $q_0$.
Conversely, the errors that do not affect $q_0$ in this ``STATE PREP" section will not be detected by $a_1$ but the likelihood of such errors is low.

\par
To verify whether the detection of error events by $a_1$ indeed enables detection of multi-qubit errors, we study the preparation of the input logical state, $|\overline{00}\rangle$, as shown in Figure \ref{fig:errorcirc} under the depolarizing noise channel.  
We calculate the contribution towards improvement in fidelity of each post-selection method at this initial stage, when errors detected by each method are for the same final state, $|\overline{00}\rangle$, as opposed to the circuit shown in Figure \ref{fig:enccirc} where the PSA method is used to detect errors during initial input state preparation and PSP at the end of the circuit. We run density matrix simulations of the circuit shown in Figure \ref{fig:errorcirc} for preparing the encoded logical $|\overline{00}\rangle$ state $|\phi\rangle$ shown in Equation \ref{eq:sprep} with increasing depolarizing noise and use operators to project states. Since this circuit does not include the ancilla $a_2$, the projection operators differ slightly and are presented in Equation \ref{eq:spprojector}. 

\par
The key comparison to understand the contribution of ancilla, $a_1$, is between the fidelity $F_{S_{P}}$ and $F_{S_{AP}}$ of states projected by $S_P$ and $S_{AP}$, respectively. Once the single qubit bit-flip errors are projected out of the prepared logical $|\overline{00}\rangle$ state, using $S_P$, any improvement in fidelity due to projection by $S_{AP}$ is entirely due to the contribution of $a_1$ and due to detection of multi-qubit error events.

\
\vspace{-10mm}
\subsection{Realistic Hardware Noise}
In addition to the study of depolarizing gate noise, we perform a similar analysis of accuracy using a realistic quantum computing hardware model. We demonstrate evaluation of the computational accuracy of a commercial device using the framework described in the previous section using the Quantinuum device emulator. We estimate the energy at the analytically determined optimal parameter with the emulator for the Quantinuum H1-1 device. The Quantinuum H1-1 device is a trapped-ion quantum charge coupled device constructed with $\ce{^{171}Yb+}$ ions as qubits and a "race-track" architecture with zones that have focused laser beams to implement gates \cite{ryan-anderson_implementing_2022}. The ions are transported to these zones for operations and the number of zones indicates the number of parallel operations possible in the device. The H1-1 device has five such zones, 20 qubits and all-to-all connectivity. The maximum number of shots that are executed for a single circuit is $10000$. We accordingly perform multiple executions of $10000$ shots each to reach the targeted $188000$ shots and we combine the resulting measurements before estimating the results. As described in the product data sheet for the Quantinuum H1-1 emulators, the emulator is designed to closely mimic the noise profile of the device. At the time of our simulations, the version of the latest product data sheet of the emulator was 6.8.3 and was dated 18th Jul 2024. The physical noise parameters listed in this version of the data sheet are reproduced in  Table \ref{tab:H11E_noise_params} and are the parameters for the noisy simulations in this study. 
\par
The Single-qubit and Two-qubit Fault Probability are the single- and two-qubit gate noise, respectively, and are largely modeled with asymmetric depolarizing noise channels.
\begin{table*}
    \centering
    \caption{Physical noise parameters reproduced from version 6.8.3 of the product data sheet of the Quantinuum H1-1E emulator.}
    \begin{tabular}[c]{|m{9cm}|m{3.5cm}|}
        \hline
        \textbf{Noise Parameter} & \textbf{Noise Parameter Value}\\
        \hline
         Single-qubit Fault Probability (p1) & $2.1\times10^{-5}$ \\
         \hline
         Two-qubit Fault Probability (p2) &  $8.8\times10^{-4}$\\
         \hline
         Bit Flip Measurement Probability (0 outcome) & $1.0\times10^{-3}$\\
         \hline
         Bit Flip Measurement Probability (1 outcome) & $4.0\times10^{-3}$\\
         \hline
         Crosstalk Measurement Fault Probability & $1.45 \times 10^{-5}$\\
         \hline
         Initialization Fault Probability & $3.62\times10^{-5}$\\
         \hline
         Crosstalk Initialization Probability & $5.020\times10^{-6}$\\
         \hline
         Ratio of Single-Qubit Spontaneous Emission to p1 & $0.54$\\
         \hline
         Ratio of Single-Qubit Spontaneous Emission in Two-Qubit Gate to p2 & $0.43$\\
         \hline
    \end{tabular}
    \label{tab:H11E_noise_params}
\end{table*}
\noindent
 Fractions of Single- and Two-qubit gate noise that represents spontaneous emission instead of the asymmetric depolarizing noise are represented by Ratio of Single-Qubit Spontaneous Emission to p1 and Ratio of Single-Qubit Spontaneous Emission in Two-Qubit Gate to p2, respectively. The Bit Flip Measurement Probability (0 outcome) indicates probability of measuring $1$ when the qubit is in $0$ and vice versa for Bit Flip Measurement Probability (1 outcome). Initialization Fault Probability, as the name suggests, indicates probability of error at qubit initialization. Crosstalk error rates during measurement and initialization are indicated by Crosstallk Measurement Fault and Crosstalk Initialization Probability, respectively.
\par
 We estimate the energy at the optimal parameter for the unencoded and encoded ansatzes of molecular hydrogen on the emulator with their default error model with error parameters shown in Table  \ref{tab:H11E_noise_params}. For the simulation with the best energy estimate, we run simulations at three different parameters, $\theta \in [-0.400606, -0.22967,-0.058734]$, including the known optimal parameter $\theta^* = -0.22967$ to verify that the minimum energy estimate corresponds with the analytical optimal parameter.  We calculate the energy expectation values for both the unencoded and encoded simulations, and for each post-selection method, with and without RED using $N=188000$ shots. 
  \par
 For circuits with encoding for [3,1]-RED, we post-select outcomes based on unanimous vote. For each set of data and the corresponding readout qubits, $a_i,k_i, l_i$ for $i \in [1,2]$ and $q_i, k_j,l_j$ for $i \in [0,3]$ and $ j \in [1,4]$, of the [[4,2,2]]-encoded ansatz as shown in Figure \ref{fig:roed_enc}, and $q_i, r_i, s_i$ for $i \in [0,1]$ of the unencoded ansatz shown in Figure \ref{fig:roed_unenc}, we discard outcomes in which measurements of all three qubits in the encoding are not equal, as they indicate readout error has occurred on one or two of the qubits. For the [[4,2,2]]-encoded ansatz we first post-select measurement outcomes based on RED followed by the post-selection strategies listed for the [[4,2,2]]-encoding in section \ref{subsubsection:PSStrat}. As in the case for the depolarizing error model, for each post-selection method from the simulations under the H1-1E error model we calculate the probability of success ($\eta$), and standard error of the mean as described in the previous section. We execute ansatzes in parallel where possible. For instance, about nine circuits of the unencoded two-qubit hydrogen ansatz shown in Figure \ref{fig:unenccirc} can be implemented in parallel on the 20 qubit Quantinuum H1-1E device and emulator. The all-to-all connectivity proves useful for such executions. 
\par
We compare these simulated results for the expectation value of energy and accuracy against the chemical accuracy benchmark and against results from numerical simulations under the depolarizing error model. We evaluate the impact of using the [[4,2,2]] QED code on hardware using the emulator and additionally compare the results against the estimates from circuits with RED to evaluate the effectiveness of the readout error encoding against the readout noise included in the emulator noise model. We then estimate the resources required to perform these simulations and compare the results between the unencoded and [[4,2,2]]-encoded simulations.

\
\section{Results}
We first present the results for simulations under the depolarizing error model followed by those of the Quantinuum H1-1E emulator.
\subsection{Depolarizing Error Model}
\par
 The [[4,2,2]] encoded VQE circuit shown in Figure \ref{fig:enccirc} and the unencoded circuit shown in Figure \ref{fig:unenccirc} were simulated using the IBM `Aer' simulator with $N=2\times10^5$ shots under a standard depolarizing noise model. All qubits were measured and the measurement counts were used to calculate expectation values of the Hamiltonian for the unencoded circuit, and the encoded circuit before and after post-selection. 
\par
The energy expectation values, variance of the calculation and probability of success at the noise parameter value at which the energy estimate reached chemical accuracy ($p=0.09\%$) are presented in Table \ref{tab:numbers}. Without post-selection, the energy of the encoded ansatz is much higher than the unencoded ansatz and the exact energy of $-1.13712$ Ha. PSA improves the energy of the encoded ansatz while still falling short of the unencoded ansatz by $0.2\%$ and PSP leads to a lower energy than PSA and the unencoded ansatz. The combined method, PSAP, reaches chemical accuracy at the highest noise parameter value of $0.09\%$ than all other methods.  PSAP leads to the highest improvement in the accuracy of the energy estimate and brings the energy estimate within chemical accuracy of $1.6$ mHa.
\begin{table*}
    \centering
    \caption{Comparing outcomes of the numerical simulations of the unencoded and [[4,2,2]] encoded ansatzes with and without post-selection at $0.09\%$ depolarizing noise parameter value and $N=2\times10^5$ shots.}
    \begin{tabular}{|c|c|c|c|}
        \hline
         & \textbf{Energy (mHa)} & \textbf{Variance (mHa)} & \textbf{Prob. of Success (\%)} \\
         \hline
         \textbf{Unencoded} & $-1134.81 \pm 0.52$ & 5.38 & 100\\
         \hline
         \textbf{Encoded No PS} & $-1131.40 \pm 0.72$ & 5.10 & 100\\
         \hline
         \textbf{PSA} & $ -1132.88 \pm 0.71$ & 4.99 & $99.644 \pm 0.002$ \\
         \hline
         \textbf{PSP} & $-1134.87\pm 0.70$ & 4.84 &$99.472 \pm 0.002$\\
         \hline
         \textbf{PSAP} & $-1135.89\pm0.69$ & 4.76 & $99.232 \pm 0.003$\\
         \hline
    \end{tabular}
    \label{tab:numbers}
\end{table*}
\par
\begin{figure}
    \centering
    \includegraphics[width=0.85\linewidth]{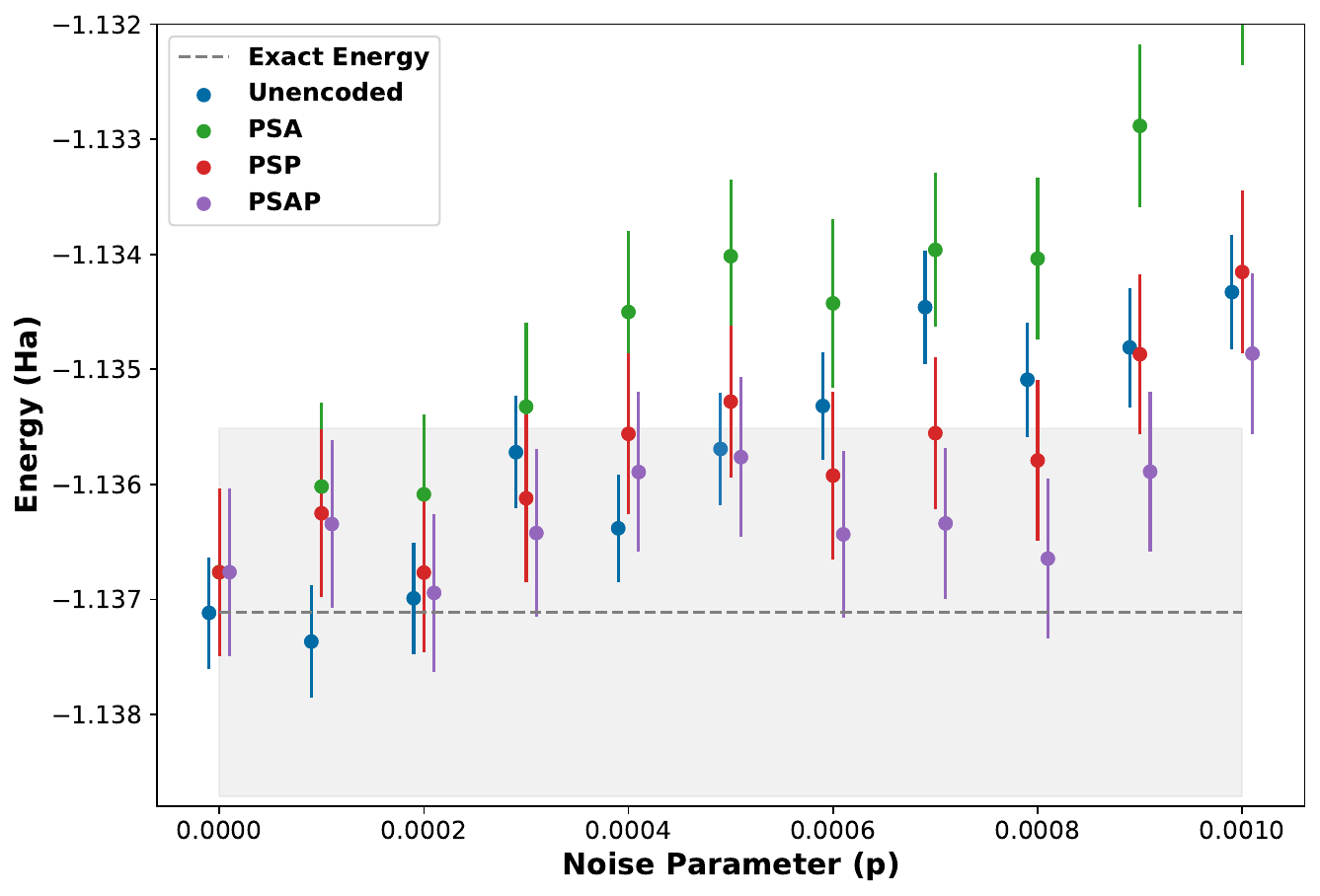}
    \caption{Estimated energy of post-selected outcomes and unencoded simulation compared against the benchmark of chemical accuracy. Shaded region represents the region of chemical accuracy and is set to $\pm 1.6$ mHa}
    \label{fig:chemaccenergy}
\end{figure}
\begin{figure}
    \centering
    \includegraphics[width=
    0.8\linewidth]{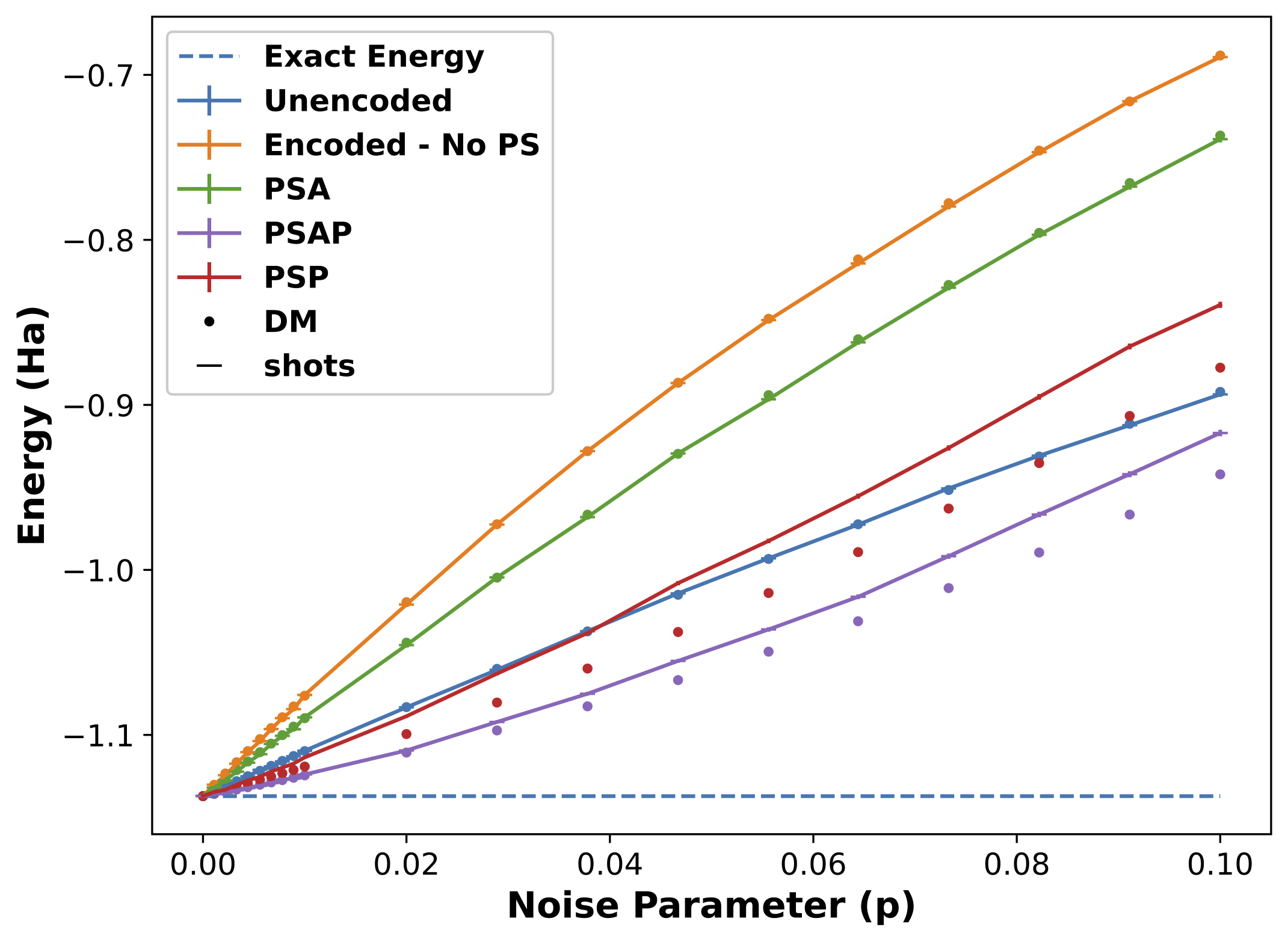}
    \caption{Energy expectation values calculated from numerical simulations of the unencoded and [[4,2,2]] encoded ansatz without (No PS) and with post-selection methods, PSA, PSP and PSAP with increasing two-qubit gate noise at 200000 shots using the IBM aer simulator and a standard depolarizing noise model.}
    \label{fig:shotsenergy}
\end{figure}

The energy estimates at the lower end of the noise range considered are compared against the benchmark of chemical accuracy in Figure \ref{fig:chemaccenergy} for all post-selection methods. The trends are similar to those observed in Table \ref{tab:numbers}. The PSAP method is within chemical accuracy at noise parameters $\leq 0.09\%$ followed closely by the PSP method, which reaches chemical accuracy at $\leq 0.08\%$ noise. The trend continues even at larger noise values as shown in Figure \ref{fig:shotsenergy}. PSA improves the energy estimate over encoded ansatz simulation with no post-selection. However, it results in higher energy than the unencoded ansatz. PSP results in lower energy than both the encoded ansatz without post-selection and PSA. The results are also lower than the unencoded ansatz below noise level of 4\%. Combining the two post-selection methods, PSAP results in the lowest energy and for all noise levels considered. Error bars in this plot of Figure \ref{fig:shotsenergy} are too small to be visible and are in the range $0.06-2.8$mHa.
\par
We also present expectation value calculations using the exact density matrix for all simulations with increasing two-qubit gate noise in Figure \ref{fig:shotsenergy}. Since calculations using the exact density matrix represent estimates of the energy in the limit of infinite shots, the energy estimate from the simulations with finite number of shots should agree with the exact density matrix calculations. They are in agreement for all simulations except the calculations for the PSP and PSAP method. For both methods, energy estimated from simulations with shots is higher than calculations using the exact density matrix.

\par
\par
\begin{figure}
    \centering
        \includegraphics[width =0.8\linewidth]{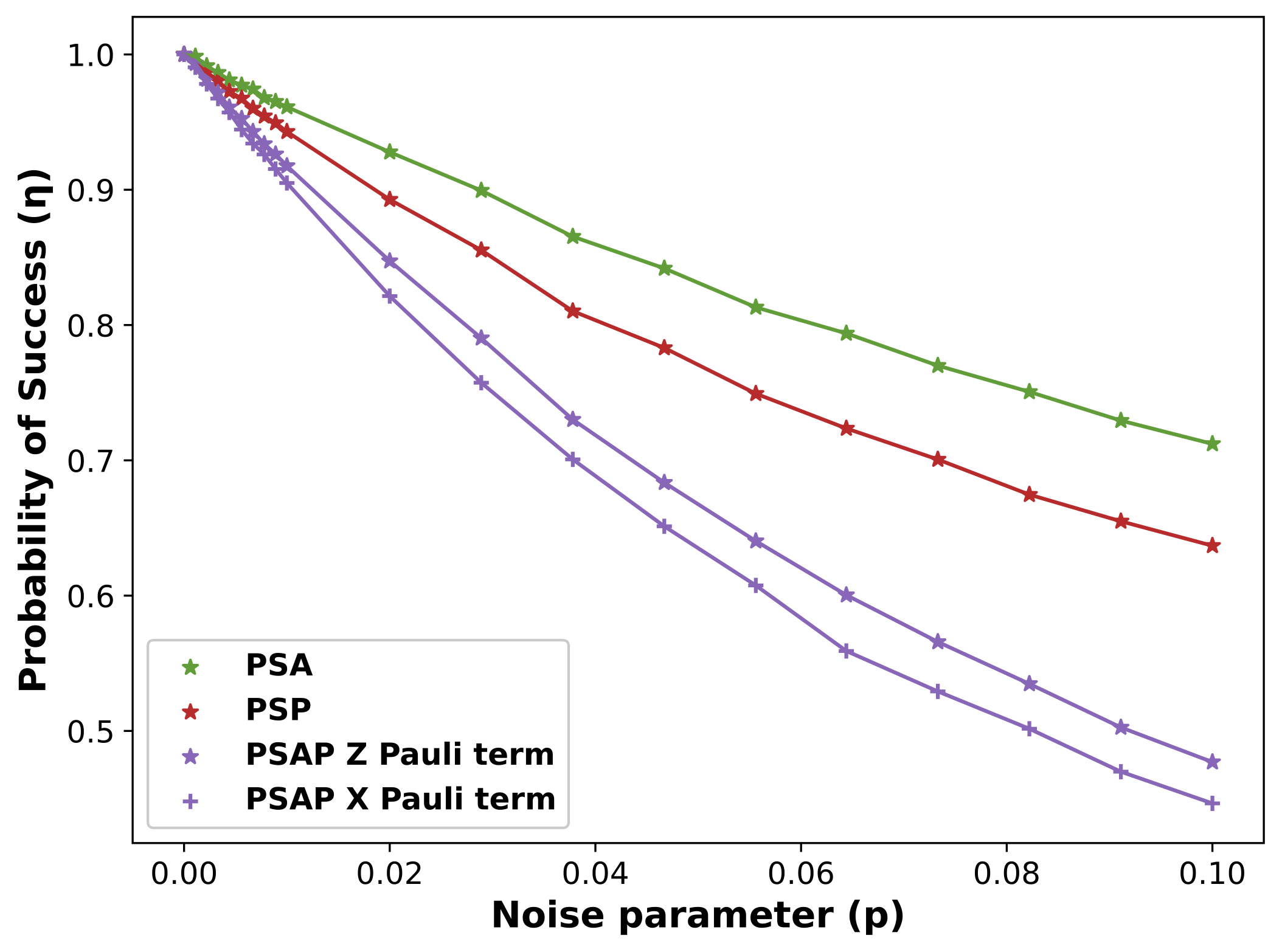}
    \caption{Probability of success of each post-selection method, PSA, PSP and PSAP of the [[4,2,2]] encoded ansatz under a standard depolarizing model. PSAP X and Z Pauli term plots are representative of the post-selection outcomes for circuits with and without the Hadamard rotation applied prior to measurement on all the encoded qubits ($q_0-q_3$) in Figure \ref{fig:enccirc}, respectively. Trends between the two Pauli terms are similar for all post-selection methods.}
    \label{fig:PofSuccess}
\end{figure}

The consequence of post-selection is having fewer samples for calculating energy expectation values. In all cases, including the noiseless case, since we are only considering half of the samples from the uniform superposition of the final state due to non-transversal rotation by ancilla $a_2$, i.e., only samples with $a_2 = |0\rangle$, we start with $\sim 50\%$ of the original number of shots prior to implementation of any post-selection method due to detection of errors. We present the change in probability of success after each post-selection method with increasing noise in Figure \ref{fig:PofSuccess}, and in Table \ref{tab:numbers}, for a specific noise parameter value, normalized after selecting samples with $a_2 = |0\rangle$. 
\par
For all post-selection methods, the probability of success decreases with increasing noise indicating an increasing proportion of states with detected errors. This proportion is also determined by the method of post-selection. PSA retains the highest proportion of samples. However, energy expectation value calculations indicate that this does not lead to an improvement over the unencoded ansatz. Post-selection by logical state parity, PSP, results in a lower probability of success than PSA, while the highest proportion of samples are discarded due to the combined post-selection method, PSAP. Standard error of the mean in the figure is too small in magnitude to be visible but ranges from 0.006\% to 0.2\%. There is also a difference in the probability of success depending on the Pauli term being measured as shown in Figure \ref{fig:PofSuccess}. Probability of success for $X$ Pauli term measurements are slightly lower than for $Z$ Pauli term measurements due to additional noise in the circuit from the Hadamard gates used to rotate the state prior to making the measurement. The PSAP $X$ and $Z$ Pauli term plots are representative and apply to all other post-selection methods.
\par
The number of samples retained consequently impact the precision of the calculation. Table \ref{tab:numbers} shows that there is a decrease in precision/ increase in SEM of the energy estimate from the encoded simulations with and without post-selection compared to the unencoded simulation. However, the variance of the calculation for the encoded post-selected calculations are lower than the unencoded simulations and lowest for PSAP calculation, which also results in the highest accuracy of the energy estimate. The decrease in the SEM, therefore, is a result of considering only half as many samples in the calculation for the encoded ansatz as the unencoded ansatz, where the full $N=2\times10^5$ samples are included in the SEM calculation.
\par
\begin{figure}
    \centering
        \includegraphics[width =0.8\linewidth]{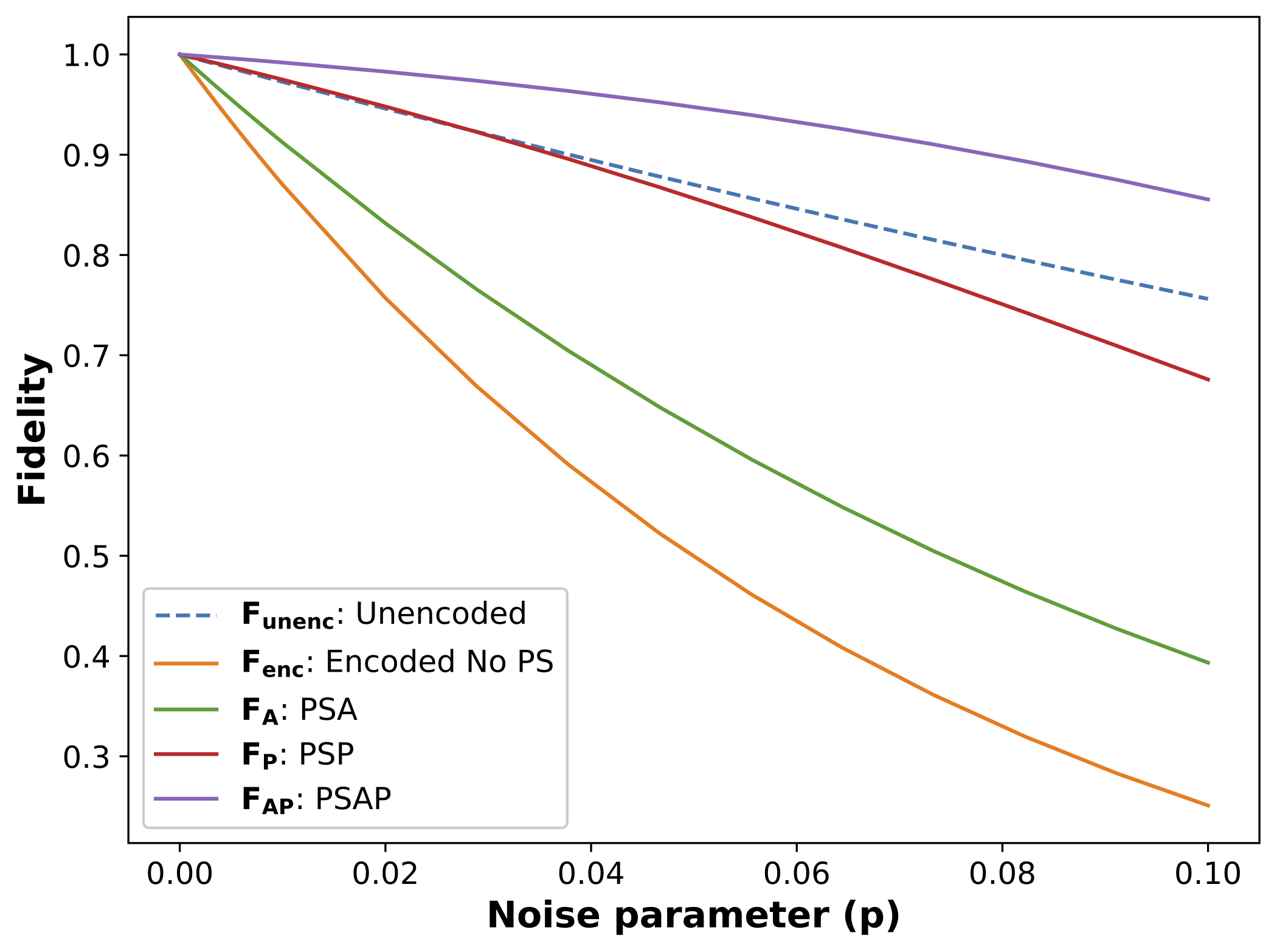}    
    \caption{Fidelity of the states prepared by numerically simulating the respective unencoded state, [[4,2,2]] encoded state and projected states corresponding to the post-selection methods, PSA, PSP and PSAP, with increasing standard depolarizing noise, $p$.}
    \label{fig:fidelity}
\end{figure}

We now present the logical fidelity of the the unencoded state, and the encoded state, without and with projections $\Pi_i \in \{\Pi_A,\Pi_P, \Pi_{AP}\}$ corresponding with post-selection methods PSA, PSP and PSAP, respectively, with increasing two-qubit depolarizing noise $p$ in Figure \ref{fig:fidelity}. The logical fidelity decreases with increasing noise, and changes with each post-selection strategy. The state fidelity of the unencoded ansatz $F_{unenc}$, decreases with noise but is consistently better than the encoded state fidelity $F_{enc}$ even after applying $\Pi_A$, as indicated by $F_A$.
A drastic improvement in fidelity is observed for states projected using projector $\Pi_P$. Not only is this fidelity, $F_P$, higher than $F_A$, it outperforms $F_{unenc}$ up to a noise value of $\sim2\%$. Combining the two post-selection methods results in the best fidelity, $F_{AP}$, for this circuit and is better than $F_{unenc}$ for all noise parameters considered.
Furthermore, overall decrease in fidelity of all cases with increasing noise in spite of projecting states against single- and multi-qubit errors is due to errors that go undetected in each projection, such as logical errors, introduced by the UCC ansatz or by combination of errors from state-preparation and UCC ansatz, that escaped detection by projection using $\Pi_{AP}$.  
\begin{figure}
    \centering
    \includegraphics[width=0.8\linewidth]{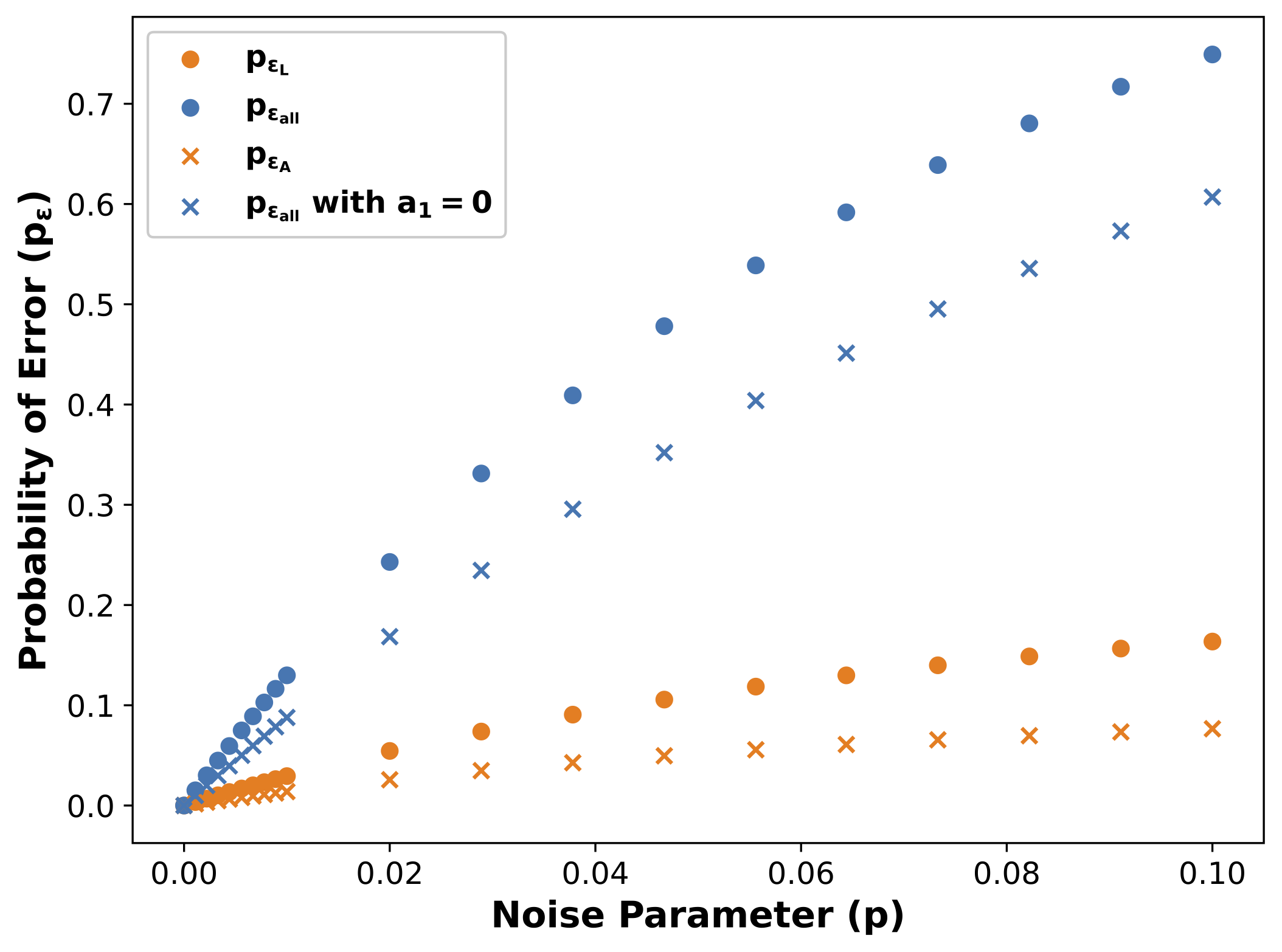}
    \caption{Probability of Error, ($p_{\epsilon}$) in the encoded  circuit with increasing two-qubit depolarizing noise. Plot labeled with $a_1=0$ are errors in the state projected with $a_1=0$}
    \label{fig:LogError}
\end{figure}
\par
Both fidelity and accuracy of the energy estimates in the simulated encoded and unencoded circuits are impacted by the errors introduced by the depolarizing error model. In the encoded circuit, in particular, the single- and two-qubit gate errors in the noise model manifest as logical, $p_{\epsilon_{L}}$ or non-logical errors ($p_{\epsilon_{NL}}$) in the final noisy mixed state. We present the probability of logical errors, $p_{\epsilon_{L}}$, in the encoded circuit and in the projected state with $a_1=0$ ($p_{\epsilon_{A}}$) in Figure \ref{fig:LogError} and also present the probability of all errors with increasing two-qubit depolarizing noise. While the errors increase with increasing noise, the proportion of all errors that are logical is much smaller. Since a logical error in this encoding requires an error event on a minimum of two qubits, the small proportion of logical errors indicate that a large proportion of errors in the encoded circuit are non-logical, single- and multi-qubit errors. Both $p_{\epsilon_{all}}$ and $p_{\epsilon_{A}}$ for states projected with $a_1=0$ are lower than the original state. 
\par
\begin{figure}
    \centering
    \includegraphics[width = 0.8\linewidth]{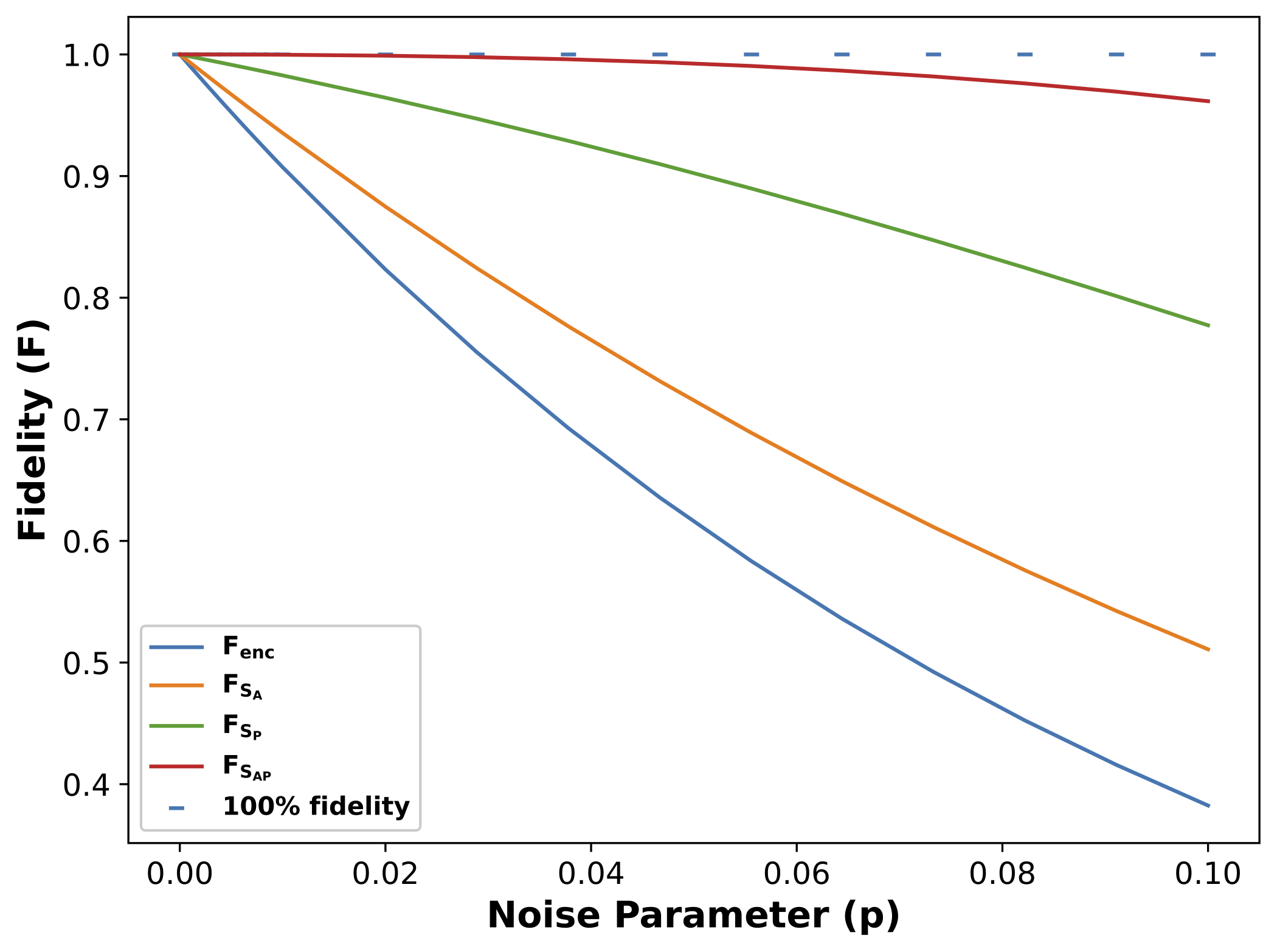}
    \caption{State fidelity for the noisy logical input $\overline{|00\rangle}$ state preparation and the projected states from the circuit labeled ``STATE PREP'' in Figure \ref{fig:enccirc} and shown in Figure \ref{fig:errorcirc}}
    \label{fig:erroranalysis}
\end{figure}

We also present the fidelity with increasing two-qubit depolarizing noise $p$ for the encoded input state $|\overline{00}\rangle$ and the states projected using $S_A, S_P$ and $S_{AP}$ in Figure \ref{fig:erroranalysis} to better understand the contribution of $a_1$ to the improvement in the accuracy of the energy estimate. We compare the fidelity of the state projected by $S_P$ with $S_{AP}$ for the same final encoded input state $|\overline{00}\rangle$. Improvement in fidelity, $F_{S_P}$, is due to the removal of states with single qubit bit-flip errors by projection operator $S_P$. Combining the methods using projector $S_{AP}$ not only improves the fidelity over the fidelity $F_{S_P}$ for the state projected by $S_P$, but also results in fidelity that is nearly $1$ at $\lesssim 5\%$ noise.

\subsection{Accuracy under a Realistic Noise Model}
We present results from noisy simulations on the Quantinuum H1-1E emulator with their default error model. Their default error model closely mimics their H1-1 device. We present energy estimates and the corresponding SEM for molecular hydrogen with the unencoded ansatz and the encoded ansatz with and without post-selection. For results with post-selection, we also show the POS. We additionally present these results with the encoding for RED for both the unencoded and encoded ansatzes with the corresponding the probability of success. We describe the reference names used for the different simulations presented in this section in  Table \ref{tab:sim_disc}.
\begin{table*}
    \centering
    \caption{Index of reference names used for different simulations with their descriptions.}
    \begin{tabular}[c]{|m{30ex}|m{7cm}|}
       \hline
       \textbf{Simulation Reference}  & \textbf{Description}\\
       \hline
       Unenc\slash Enc noiseless & Noiseless simulation of the unencoded/encoded ansatz on the Quantinuum H1-1 emulator.\\
       \hline
       Unencoded & Simulation of unencoded ansatz with default error model of the Quantinuum H1-1 emulator.\\
       \hline
       Unenc [3,1]-RED & Simulation of unencoded ansatz with [3,1]-RED encoding with default error model of the Quantinuum H1-1 emulator.\\
       \hline
       Enc PSA\slash PSP\slash PSAP\slash no PS & Simulation of [[4,2,2]]-encoded ansatz with default error model of the Quantinuum H1-1 emulator with PSA, PSP, PSAP or without PS.\\
       \hline
       Enc [3,1]-RED \newline PSA\slash PSP\slash PSAP\slash no PS & Simulation of [[4,2,2]]-encoded ansatz with [3,1]-RED encoding on the Quantinuum H1-1 emulator with default error model with PSA, PSP, PSAP or without PS.\\
       \hline 
    \end{tabular}
    \label{tab:sim_disc}
\end{table*}
\par
We first present results of the energy estimates of simulation of the unencoded ansatz and the encoded ansatz with post-selection with and without RED in  Table \ref{tab:rod_sim_deets}. The table includes the energy estimates with and without RED of the unencoded ansatz and the best post-selection method of the encoded ansatz, PSAP, along with the corresponding difference, $\Delta E$, from the exact energy of $-1.13712\text{Ha}$ and the probability of success, $\eta$. 
\begin{table*}
    \centering
    \caption{Energy and Probability of Success ($\eta$) with and without RED for unencoded and [[4,2,2]]-Encoded ansatz with post-selection methods.}
    \begin{tabular}{|m{5cm}|m{2.8cm}|m{2cm}|m{1.5cm}|}
        \hline
         Simulation & Energy (mHa)  & $\Delta E$ (mHa) & $\eta$ (\%) \\
         \hline
         Unencoded  & $-1133.39\pm 0.51$  & $3.73$ & $100$\\
         \hline
        Encoded PSAP & $-1129.78\pm 0.75$ & $7.33$ & $49.2$ \\
         \hline
         Unencoded with [3,1]-RED  & $-1133.94\pm 0.51$ & $3.17$ &$98.2$ \\
         \hline
        Encoded PSAP with [3,1]-RED & $-1135.74\pm 0.73$ & $1.37$ & $47.3$ \\
         \hline
    \end{tabular}
    \label{tab:rod_sim_deets}
\end{table*}
 When estimating energy of the the [[4,2,2]]-encoded ansatz with the [3,1]-RED at three parameter values $\theta \in [-0.400606, -0.22967,-0.058734]$, $\theta^{*}$ was found to correspond with the value calculated analytically at $-0.22967$ rads.
\par
\begin{figure}
    \centering
    \includegraphics[width=0.9\linewidth]{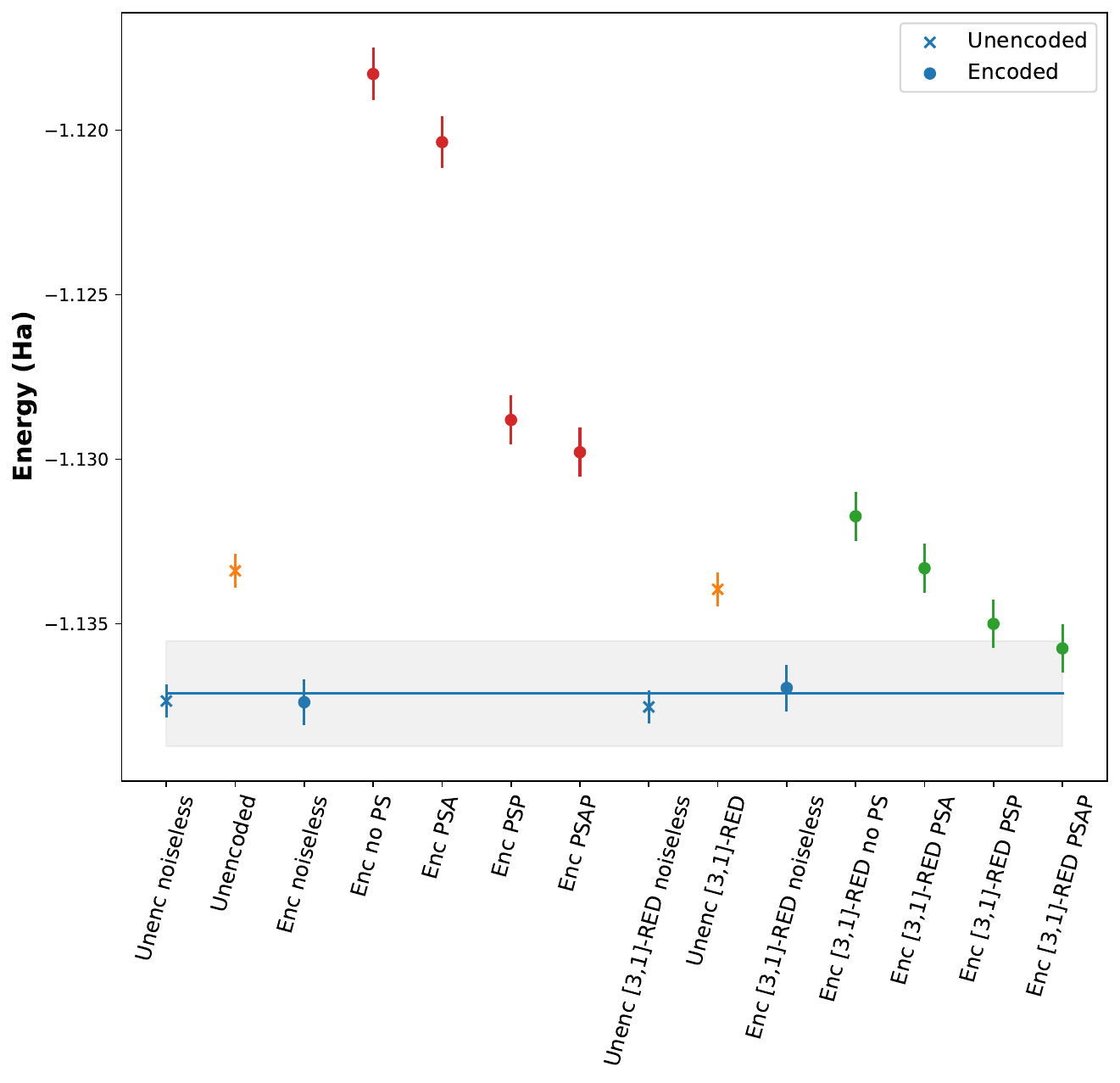}
    \caption{Energy estimates of unencoded and [[4,2,2]]-encoded simulations with post-selection without and with [3,1]-RED on the Quantinuum H1-1E emulator with the default error model. Shaded grey area represents the chemical accuracy region.}
    \label{fig:31ROD}
\end{figure}

We now show results of the energy estimates of simulations without and with RED of the unencoded ansatz, and the encoded ansatz with post-selection in Figure \ref{fig:31ROD}. This figure additionally has the corresponding noiseless estimates from the emulator as a reference. The shaded gray area represents the region within chemical accuracy or $\pm 1.6 \ \text{mHa}$ of the exact energy for molecular hydrogen at $0.74$\r{A}. Error bars indicate the standard error of the mean. In this simulation, without RED, the unencoded ansatz results in the lowest energy estimate compared to the best and lowest energy estimates from encoded ansatz, i.e., with the PSAP strategy. Neither estimate, however, reaches chemical accuracy.
\par
Employing RED leads to dramatic decrease in energy estimates for the encoded ansatz, quantitatively presented in  Table \ref{tab:rod_sim_deets}. For the unencoded ansatz, the code only results in a slight decrease in the energy compared to the results without RED. The magnitude of reduction in the energy estimate for all post-selection methods of the [[4,2,2]]-encoded ansatz when using RED is greater than the corresponding decrease in energy for the unencoded ansatz. Comparing the energy estimates with RED for the unencoded and [[4,2,2]]-encoded ansatz shows that not only does RED significantly improve the energy estimate for the [[4,2,2]]-encoded ansatz, the best estimate, the PSAP method, with the [3,1]-RED is about $1.37 \ \text{mHa}$ from the exact energy as shown in  Table \ref{tab:rod_sim_deets} and within chemical accuracy. The second best estimate, from the PSP method, is also lower than that of the unencoded ansatz. In the case of the unencoded ansatz with RED the estimates are above the chemical accuracy regime with the energy difference from the exact energy being ${\sim}3.17 \ \text{mHa}$, which is about ${\sim}1.6 \ \text{mHa}$ higher than chemical accuracy. The difference in the best estimate between the unencoded and [[4,2,2]]-encoded ansatz is about $1.8 \ \text{mHa}$.

\begin{figure}
    \centering    \includegraphics[width=0.8\linewidth]{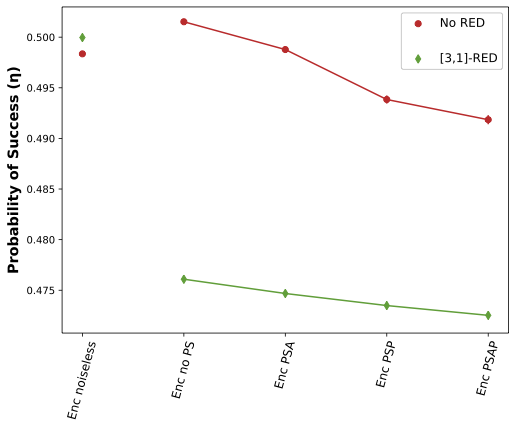}
    \caption{Probability of Success for Encoded Simulations on the Quantitnuum H1-1E emulator with and without RED.}
    \label{fig:POS_ROD}
\end{figure}

We present the probability of success for the encoded ansatz simulation in Figure \ref{fig:POS_ROD} with and without RED. There is a drop in probability of success of ${\sim}2\%$ for the post-selection methods with [3,1]-RED from the simulations without RED but the corresponding drop in energy for the method with the best estimate, the PSAP method, with [3,1]-RED is nearly $6 \ \text{mHa}$. The corresponding difference in energy for the best estimate with the unencoded ansatz is $0.6 \ \text{mHa}$ for a loss of $1.8\%$ of samples as shown in  Table \ref{tab:rod_sim_deets}. Additionally, the probability of success in the case of the unencoded ansatz is from $100\%$ and for the [[4,2,2]]-encoded ansatz, from ${\sim} 50\%$ as the implementation of gate teleportation with ancilla $a_2$ results in a uniform superposition of two equivalent states of which we only use one. 
\par
Finally, we present the resources required for the simulations of the [[4,2,2]]-encoded and unencoded ansatz for the best energy estimate with [3,1]-RED in Table \ref{tab:res_est} based on the simulations on the emulator. Shots or number of samples are displayed as twice the amount since two circuit executions are required, one each for measuring the $Z_0Z_1$ and $X_0X_1$ Pauli term. Per run indicates the number of said resource that is required per execution of the circuit inlcuding parallel circuits. Parallel circuits per run are $0$ for the encoded ansatz with [3,1]-RED as only one 18-qubit circuit can be executed on the 20-qubit emulator at a time, whereas three unencoded circuits can be executed with [3,1]-RED. Maximum number of single qubit gates point to the circuit for measuring ``$X_0X_1$" Pauli term, which requires execution of additional Hadamard gates. Both the total number of qubits required for parallel executions and number of qubits required per ansatz execution is presented.
\par
The expected cost is an approximate estimate of the total cost required as calculated by the Quantinuum emulator for executing these simulations on the Quantinuum device. Although this is not typically included in resource estimates, as devices become available for users commercially, there is a corresponding charge for accessing the devices which imposes an additional constraint that must be
\begin{table*}
    \centering
    \caption{Resource analysis for estimating the ground state energy within chemical accuracy on the Quantinuum H1-1E emulator with [3,1]-RED for the unencoded and [[4,2,2]]-encoded ansatzes.}
    \begin{tabular}{|c|c|c|}
        \hline
         \textbf{Resources}   & \textbf{Unencoded} & \textbf{[[4,2,2]]-Encoded}  \\
         \hline
         Shots & $2\times62700$ & $2\times188000$ \\
         \hline
         Parallel Circuits per Run & 3 & 0 \\
         \hline
         Number of qubits per Ansatz with [3,1]-RED & 6 & 18 \\
         \hline
         Number of qubits per Run with [3,1]-RED & 18 & 18 \\
         \hline
         Max. Single-qubit Gates per Run & 9 & 7 \\
         \hline
         Two-qubit Gates per Run & 18 & 25 \\
         \hline
         Measurements per Run & 18 & 18 \\
         \hline
         Estimated Total Runtime & 8.5 hrs & 1.5 days\\
         \hline
         Expected cost of credits & 10200 & 31400 \\
         \hline
    \end{tabular}
    \label{tab:res_est}
\end{table*}
considered when evaluating the cost of implementing such benchmarks on commercial devices, such as, Quantinuum. The cost of credits is calculated by the emulator using the formula shown in Equation \ref{eq:credqtuum} and is provided by Quantinuum in their product data sheet.
\begin{equation}
    HQC = 5 +\frac{N_{1q}+10N_{2q}+5{N_m}}{5000}C
    \label{eq:credqtuum}
\end{equation}
where $N_{1q}, N_{2q}$ and $N_m$ are number of native single-qubit gates, two-qubit gates and state preparation and measurement operations in the circuit, respectively and $C$ is the number of shots. The product data sheet also suggests that the estimated run time for circuits is about an hour for circuits that require between 500 to 1200 HQCs, which depends on different factors such as the number of the connectivity required for the circuit and the dynamic calibrations of the system. The estimated total runtime in Table \ref{tab:res_est} have been calculated by assuming execution of circuits worth 1200 credits per hour for the simple unencoded ansatz and about 800 credits per hour, the average of the two extremes of 500 and 1200, for the slightly more complicated encoded circuit.

\section{Discussion}
\par
The results indicate that using error detection for mitigating the dominant sources of error can enable the estimates to reach the targeted level of accuracy and precision, which, in this case, is chemical accuracy. Simulations with only the depolarizing error channel indicate that the [[4,2,2]] QED code improves energy estimates over the unencoded ansatz when using post-selection by logical state parity, PSP, by itself at low noise values, and in combination with post-selection by ancilla, $a_1$, measurement, PSAP, at least up to $10\%$ noise. The latter also results in energy estimates within chemical accuracy at the highest noise threshold of all the simulations. 

\par
 The decrease in precision of the estimated energy calculated from post-selected results of the simulation of the encoded ansatz compared to the unencoded ansatz is due to consideration of only one half of the results from the use of gate teleportation. This is particularly remarkable when considering that the number of samples used to calculate the energy estimate is nearly half as many as that used for the unencoded ansatz but results in only ${\sim}0.2 \ \text{mHa}$ increase in standard error with a significant improvement in accuracy. Within the different post-selection methods, improvement in precision is consistent with improvement in logical fidelity, indicating that in addition to improvement in the accuracy of the energy estimate, post-selection also improves the corresponding precision of the calculation. The improvement in precision in spite of increasing loss of samples with increasing accuracy of the post-selection method is a result of the improvement in variance of the calculation. We see similar improvement in precision for the encoded ansatz when using the RED code under the Quantituum H1-1E error model. 
 \par
Simulations under the device error model without RED indicate that the unencoded energy estimate results in a lower energy than the best post-selection strategy for the [[4,2,2]]-encoded ansatz, the PSAP energy estimate. Both ansatzes result in lower energies with RED. Using RED in addition to the PSAP method of the [[4,2,2]]-encoded ansatz results in an estimate of energy that is nearly $1.8 \ \text{mHa}$ lower than the best unencoded energy estimate and within chemical accuracy.  Numerical simulations of the [[4,2,2]]-encoded ansatz under an isotropic two-qubit depolarizing error model without any readout error indicated that the threshold at which the energy estimate reached chemical accuracy was $0.09\%$ as shown in Figure \ref{fig:chemaccenergy}. The Quantinuum emulator error model not only includes read-out error but also models depolarizing error with an asymmetric channel and includes leakage by spontaneous emission as part of the depolarizing error model. The energy estimate reaching chemical accuracy with the [3,1]-RED suggests that the readout error was nearly completely eliminated and that the [[4,2,2]]-encoding is at least as effective at detecting errors with this depolarizing error model as with the isotropic error model. Additionally, even though the other error sources listed in  Table \ref{tab:H11E_noise_params} are an order(s) of magnitude lower than the leading error sources, they were not considered in the numerical simulations with the isotropic depolarizing error model and yet resulted in a similar estimate indicating that they did not have as much of an effect on decreasing the accuracy. 
\par
The deviation of the estimated energy using the PSP and PSAP methods from the corresponding energy expectation value calculation using the density matrix, occurs due to the difference in order of operations between the two methods. In the latter, the states of interest are projected and the expectation value is calculated from these states. In the former, the bitstrings are post-selected from the measurement of the final state and then the expectation values are calculated from these post-selected bitstrings. Post-selection strategies using $a_1$ lead to improvement in accuracy of the energy estimate due to detection of single and multi-qubit error events. Nearly unity fidelity for projecting states with logical state parity and $a_1=0$ measurement ($F_{S_{AP}}$) indicates that $a_1$ detects nearly all multi-qubit error events that are left undetected by projecting states with logical state parity alone.  Additionally, the reduction in probability of logical errors in the state projected for $a_1=0$ measurement, further confirm the detection of errors other than single qubit errors by $a_1$.

\par
As an application benchmark, the study demonstrates that effectively removing errors in every sample measurement using methods such as QED and readout error detection can sufficiently manage dominant sources of errors typically found in current commercial devices and lead to the targeted accuracy if the error rate is sufficiently low. However, as can be seen from the resource analysis, the benchmark with error detection leads to a large increase in resource overhead with a three-fold increase in the number of qubits and two-qubit gates alone compared to the unencoded ansatz, not to mention the corresponding increase in the cost to use the device. Such an increase in resources suggests that scaling up the benchmark exactly as it is to systems of larger scale may be prohibitive.  
\par
Resource scaling depends both on the choice of ansatz and the quantum error detection or correction scheme. The analysis is presented for the smallest molecular system using the smallest error detection code for a device with all-to-all connectivity, making it the best case scenario in terms of resources required. Increase in size of the molecular system, constraints in device connectivity, and change in quantum error correction/detection code will inevitably lead to an increase in the resources required. While the study demonstrates that the methods lead to an effective benchmark, in order for such benchmarks to be applicable for more useful larger scale systems across device architectures innovative methods to optimize the scaling of resources with increase in system size are needed especially as technologies grow and error rates become lower. In this study itself, implementation of the rotation gate in the [[4,2,2]]-encoded ansatz without the consequence of losing nearly half the samples in the process would have led to a decrease in at least one of the resources: number of samples or shots compared to the unencoded ansatz. Similarly, optimizing ansatzes, using flag qubits to reduce overhead of larger QEC codes, and combining other error mitigation techniques may be potential avenues for future work to improve the resource overhead of the benchmark presented here with increase in system size.

\section{Conclusion}
We have demonstrated that error detection codes with post-selection lead to an energy estimate within chemical accuracy for a small two-electron system. We have presented results from noisy simulations of the ground state of molecular hydrogen using VQE under the depolarizing error model and the error model of a commercial quantum computing device, Quantinuum H1-1. The aim was to assess the impact of different post-selection methods on the energy accuracy and evaluate the outcomes against the benchmark of chemical accuracy, defined as the estimated energy being within $1.6  \ \text{mHa}$ of the exact energy, by using the [[4,2,2]] QED code and RED with post-selection to mitigate gate depolarizing error and readout error, respectively. We found that the results demonstrate the improvement in the accuracy and precision of energy estimates with the best estimate reaching chemical accuracy indicating effective mitigation of both gate and readout error. The two-qubit gate error rate of the emulator at $0.088\%$ is slightly lower than the threshold, assessed by numerical simulations with the depolarizing error model, at which the [[4,2,2]]-encoded ansatz resulted in an energy estimate within chemical accuracy for molecular hydrogen. This indicates that the readout error was sufficiently removed with the [3,1]-RED code, which was the only method to result in energy estimates within chemical accuracy.

Using [[4,2,2]]-QED with post-selection based on detection of single qubit errors alone improves the accuracy of the energy estimate at low noise levels. Combining this post-selection method with post-selection using an ancilla for single and multi-qubit error detection during input state preparation further improves the accuracy of the estimate at noise levels as high as 10\%. Constructing the input state-preparation circuit such that the control qubit for all the two-qubit $CNOT$ gates is $q_0$, which is entagled with $a_1$ makes the detection of multi-qubit errors possible.
\par
Our framework for benchmarking is an investigation of accuracy of a small quantum application on a quantum device. As described earlier, computational accuracy of a device is characterized both by the difference in magnitude of estimated outcome from the exact outcome and the precision of the estimate. The latter is an indication of how reproducible the accuracy is on the given device. Error mitigation techniques typically are applied to the estimated average and the improvement in the accuracy often leads to an increase in variance \cite{endo2021hybrid}. This framework improves both accuracy and precision by removing detected errors from every measured sample prior to calculating the average. Using additional error detection for combating readout error maintains this structure and similarly improves accuracy without decreasing the precision of the estimate. The scaling of resources with system size is the main limiting factor of this benchmark and we defer studies to improve this scaling to future work.
\begin{acknowledgments}
This work was supported the U.S. Department of Energy Advanced Scientific Computing Research program office under the Accelerated Research for Quantum Computing program. This research used resources of the Oak Ridge Leadership Computing Facility at the Oak Ridge National Laboratory, which is supported by the Office of Science of the U.S. Department of Energy under Contract No. DE-AC05-00OR22725. D.C. acknowledges the support from the “Embedding Quantum Computing into Many-body Frameworks for Strongly Correlated Molecular and Materials Systems” project, which is funded by the U.S. Department of Energy (DOE), the Office of Science, the Office of Basic Energy Sciences, and the Division of Chemical Sciences, Geosciences, and Biosciences.
\end{acknowledgments}
\appendix
\section{\label{app:hamenc} Hamiltonian encoding} 
We start with the exact electronic Hamiltonian for molecular hydrogen in second quantized form:
\begin{equation}
    \begin{split}
        H(R) &= h_{00} a^{\dagger}_0a_0 + h_{22} a^{\dagger}_2a_2 + h_{33} a^{\dagger}_3a_3 + h_{11} a^{\dagger}_1 a_1 \\&+ h_{2002} a^{\dagger}_2a^{\dagger}_0a_0a_2 + h_{3113} a^{\dagger}_3a^{\dagger}_1a_1a_3 + h_{2112} a^{\dagger}_2a^{\dagger}_1a_1a_2 \\&+ h_{0330} a^{\dagger}_0a^{\dagger}_3a_3a_0 + (h_{2332}-h_{2323})a^{\dagger}_2a^{\dagger}_3a_3a_2 \\&+ (h_{0110} - h_{0101})a^{\dagger}_0a^{\dagger}_1a_1a_0 \\&+  h_{2103} (a^{\dagger}_2a^{\dagger}_1a_0a_3 + a^{\dagger}_3a^{\dagger}_0a_1a_2)\\&+ h_{2013}(a^{\dagger}_2a^{\dagger}_0a_1a_3 + a^{\dagger}_3a^{\dagger}_1a_0a_2) 
    \end{split}
    \label{eq:secquant}
\end{equation}
The molecular orbitals $|\psi_u\rangle$ and $|\psi_g\rangle$ are given as linear combinations of the respective atomic orbitals, leading to spin orbitals $|\Psi_i\rangle$ each with spin $\uparrow$ or $\downarrow$ \cite{shee_qubit-efficient_2022, romero_strategies_2018}, as
%
\begin{subequations}
    \begin{align}
        |\psi_g\rangle &= |\psi_1\rangle + |\psi_2\rangle\\
        |\psi_u\rangle &= |\psi_1\rangle - |\psi_2 \rangle
    \end{align}
    \label{eq:MO}
\end{subequations}
\vspace{-10mm}
\begin{subequations}
    \begin{align}
        |\Psi_0\rangle &= |\psi_g(\uparrow)\rangle\\
        |\Psi_1\rangle &= |\psi_u(\uparrow)\rangle \\
        |\Psi_2\rangle &= |\psi_g(\downarrow)\rangle\\
        |\Psi_3\rangle &=  |\psi_u(\downarrow)\rangle
    \end{align}
\end{subequations}
The possible states within the spin-singlet configuration in the fermionic basis for the hydrogen molecule and the corresponding states in the two qubit basis are presented in Table \ref{tab:ftoq}. The indices correspond with those of the molecular orbitals in Equation \ref{eq:MO}.

\begin{table}
    \centering
    \caption{Mapping of fermionic basis to qubit basis for all possible states of the hydrogen molecule with the spin-singlet configuration. Indices for the fermionic basis correspond with indices in Equation \ref{eq:MO} and indices for the qubit basis correspond with the qubits used in the circuit construction of the unitary operator.}
    \begin{tabular}{|c|c|}
        \hline
        $|\Psi_3\Psi_2\Psi_1\Psi_0\rangle$ & $|q_0q_1\rangle$  \\
        \hline
         $|0101\rangle$ & $|00\rangle$\\
         \hline
         $|0110\rangle$ & $|01\rangle$\\
         \hline
         $|1001\rangle$ & $|10\rangle$\\
         \hline
         $|1010\rangle$ & $|11\rangle$\\
     \hline
    \end{tabular}
    \label{tab:ftoq}
\end{table}
\par 
We define the qubit excitation, de-excitation and number operators as in \cite{shee_qubit-efficient_2022} 
\begin{equation}
    \begin{split}
        Q^+ &= |1\rangle\langle0| = \frac{1}{2}(X-iY)\\
        Q^- &= |0\rangle\langle1| = \frac{1}{2}(X+iY)\\
        N^{(0)} &= |0\rangle\langle0| = \frac{1}{2}(I+Z)\\
        N^{(1)} &= |1\rangle\langle1| = \frac{1}{2}(I-Z)\\
    \end{split}
    \label{eq:qubitops}
\end{equation}
where $X$, $Y$, and $Z$
are the Pauli operators. These operators are used to construct the reduced two-qubit Hamiltonian by using the mapping presented in Table \ref{tab:opmap}.

\begin{table*}
    \centering
    \caption{Fermionic (de)excitation and number operators mapped to the corresponding qubit and Pauli operators, where $E_{pq}$ represents the excitation and number operators when $p\neq q$ and $p=q$, respectively. The Pauli operators are constructed using Equation \ref{eq:qubitops}.}
    \begin{tabular}{|c|c|c|c|}
        \hline
        \textbf{Operator} & \textbf{Fermionic operators} & \textbf{Qubit operators} & \textbf{Pauli operators for qubits} \\
        \hline
        $E_{10}$ &  $|0110\rangle\langle0101|+|1010\rangle\langle1001|$ & $|01\rangle\langle00| + |11\rangle\langle10|$ & $\frac{1}{2} (X_0-iY_0)$  \\
        \hline
        $E_{32}$ & $|1001\rangle\langle0101|+|1010\rangle\langle0110|$ & $|10\rangle\langle00| + |11\rangle\langle01|$& $\frac{1}{2}(X_1-iY_1)$  \\
        \hline
        $E_{00}$ & $|0101\rangle\langle0101|+|1001\rangle\langle1001|$ & $|00\rangle\langle00| + |10\rangle\langle10|$ & $\frac{1}{2}(I+Z_0)$  \\
        \hline
        $E_{11}$ & $|0110\rangle\langle0110|+|1010\rangle\langle1010|$ & $|01\rangle\langle01| + |11\rangle\langle11|$ & $\frac{1}{2} (I-Z_0)$ \\
        \hline
        $E_{22}$ & $|0101\rangle\langle0101| + |0110\rangle\langle0110|$ & $|00\rangle\langle00| + |01\rangle\langle01|$ &  $\frac{1}{2} (I+Z_1)$\\
        \hline
        $E_{33}$ & $|1001\rangle\langle1001| + |1010\rangle\langle1010|$& $|10\rangle\langle10| + |11\rangle\langle11|$& $\frac{1}{2} (I-Z_1)$  \\
        \hline
    \end{tabular}
    \label{tab:opmap}
\end{table*}
\par
The electronic Hamiltonian
is then reduced as 
\begin{equation}
    \begin{split}
        H &= \sum_{pq} h_{pq} E_{pq} + \frac{1}{2} \sum_{pqrs} h_{pqrs} \delta_{qr}E_{ps} - E_{pr}E_{qs}\\
    \end{split}
    \label{eq:epq}
\end{equation}
where $E_{pq}= a^{\dagger}_pa_q$ includes terms with $p=q$ and are presented in Table \ref{tab:opmap}. 
Using these operators we proceed to transform the electronic Hamiltonian in Equation \ref{eq:secquant} to a qubit representation:
\begin{equation}
    \begin{split}
        H
        &=
        (h_{00}+h_{33}+\frac{1}{4}(h_{2002}+h_{3113}+h_{2112}+h_{0330})) I \\&+ (h_{00} - h_{11}+ \frac{1}{4} (h_{2002} - h_{3113} - h_{2112} +h_{0330})) Z_0 \\&+ (h_{22} - h_{33} +\frac{1}{4} (h_{2002} - h_{3113} + h_{2112} - h_{0330}))Z_1\\&+ \frac{1}{4}(h_{2002} + h_{3113} - h_{2112}-h_{0330})Z_1Z_0 + 0 + 0 \\&+
        h_{2103}(\frac{1}{2}(-X_1X_0-Y_1Y_0)\\&+
        h_{2013}(\frac{1}{2}(-X_1X_0+Y_1Y_0)\\
    \end{split}
    \label{eq:hameq}
\end{equation}
\begin{equation}
    \begin{split}
        g_0 &= h_{00}+h_{33}+\frac{1}{4}(h_{2002}+h_{3113}+h_{2112}+h_{0330})\\
        g_1 &= h_{00} - h_{11}+ \frac{1}{4} (h_{2002} - h_{3113} - h_{2112} +h_{0330})\\
        g_2 &= h_{22} - h_{33} +\frac{1}{4} (h_{2002} - h_{3113} + h_{2112} - h_{0330}) \\
        g_3 &= \frac{1}{4}(h_{2002} + h_{3113} - h_{2112}-h_{0330})\\
        g_4 &= -h_{2103}\\
    \end{split}
\end{equation}

Additionally:
\begin{equation}
    \begin{split}
        g_1 &= g_2\\
        h_{2013} &= h_{2103}\\
        h_{2112} &= h_{0330}
    \end{split}
\end{equation}

This leads to the final two-qubit Hamiltonian shown in Equation \ref{eq:Ham2}.
\section{Unitary operator}\label{app:unitop}
The unitary operator approximated to the first Trotter step is shown in Equation \ref{eq:trot} \cite{barkoutsos_quantum_2018}. 
\begin{equation}
    \begin{split}
    U(\theta) &= e^{T(\theta)-T(\theta)^{\dagger}} = e^{\sum_i \theta_i (\tau_i - \tau_i^{\dagger})}  \\ &\approx (\prod_i e^{\frac{\theta_i}{t} (\tau_i - \tau_i^{\dagger})})^t \\
    &\approx \prod_ie^{ \theta_i (\tau_i - \tau_i^{\dagger})}\\
    \end{split}
    \label{eq:trot}
\end{equation}
\begin{equation}
    \begin{split}
        T(\theta) &= \sum_k T_K(\theta)\\
        T_1 &= \sum\limits_{q \in occ, p\in virt} \theta_q^p a^{\dagger}_pa_q\\ T_2 &= \sum\limits_{r>s \in occ, p>q \in virt} \theta_{rs}^{pq} a^{\dagger}_p a^{\dagger}_q a_r a_s
    \end{split}
    \label{eq:exop}
\end{equation}
where $T_k$ is the excitation operator manifold and $\theta_q^p$ and $\theta_{rs}^{pq}$ are singles and doubles amplitudes, respectively and $p,q$ and $r,s$ are virtual or unoccupied and occupied orbitals, respectively.
To transform the fermionic unitary operator to a qubit representation specific to our encoding we use qubit operators from Equation \ref{eq:qubitops}. Considering only the doubles contribution in Equation \ref{eq:trot} and Equation \ref{eq:exop}, we replace the fermionic operators in $T_2$ with qubit operators using Equation \ref{eq:qubitops} as shown in Equation \ref{eq:unitder}.

\begin{eqnarray}
       T_2 =& \sum \limits_{p>q\in occ., r>s \in virt.} \theta_{rs}^{pq}(a^{\dagger}_pa^{\dagger}_q a_ra_s - a_sa_ra^{\dagger}_qa^{\dagger}_p)\nonumber\\
        =&\theta^{20}_{31} (a^{\dagger}_2a^{\dagger}_0a_3a_1 - a^{\dagger}_1a^{\dagger}_3a_0a_2)\nonumber \\
       =& \theta^{20}_{31}(E_{10}E_{32}-E_{23}E_{01}) \nonumber\\=& i\theta^{20}_{31}\frac{1}{2} (-Y_0X_1-Y_0X_1Z_0Z_1)
    \label{eq:unitder}
\end{eqnarray}

where we have used $Y_kZ_k = iX_k$, $X_kZ_k=-iY_k$, to factorize the operator in the final step. Considering our reference state $|00\rangle$, we arrive at the reduced UCC doubles ansatz as shown in Equation \ref{eq:unitaryd}:

\begin{eqnarray}
        U(\theta)=&e^{i\frac{\theta}{2}(-Y_0X_1)}e^{i\frac{\theta}{2}(-Y_0X_1Z_0Z_1)}\nonumber\\=&e^{i\frac{\theta}{2}(-Y_0X_1)}e^{i\frac{\theta}{2}(-Y_0X_1)}\nonumber\\=&e^{-i\theta Y_0X_1}
    \label{eq:unitaryd}
\end{eqnarray}



\section{Basis states and operations for the [[4,2,2]] code\label{app:basisst}}
The basis states for the logical codespace of this encoding are shown in Equation \ref{eq:basisstate} and the corresponding physical operations for each logical operation are described in Table \ref{tab:ops}. 

\begin{equation}\label{eq:basisstate}
\begin{split}
        \overline{|00\rangle} &= \frac{1}{\sqrt{2}} (|0000\rangle + |1111\rangle)\\
        \overline{|10\rangle} &= \frac{1}{\sqrt{2}} (|0101\rangle + |1010\rangle) \\
        \overline{|01\rangle} &= \frac{1}{\sqrt{2}} (|0011\rangle + |1100\rangle) \\
        \overline{|11\rangle} &= \frac{1}{\sqrt{2}} (|0110\rangle + |1001\rangle) \\
\end{split}    
\end{equation}
        

\begin{table}
        \centering
        \caption{Basis operations.}
        \begin{tabular}{c|c}
        \hline
        \multicolumn{2}{|c|}{Basis
        Operations}\\
        \hline
        Logical Basis & Physical Basis  \\
        \hline 
        $X_1 \otimes I_2$ & $X \otimes I \otimes X \otimes I$ \\
        $I_1 \otimes X_2$ & $X\otimes X \otimes I \otimes I$\\
        $Z_1 \otimes I_2$ & $Z \otimes Z \otimes I \otimes I$\\ 
        $I_1 \otimes Z_2$ & $Z \otimes I \otimes Z \otimes I$\\
        $H_1 \otimes H_2 $ & $H \otimes H \otimes H \otimes H$ \\
        $CNOT_{12}$ & $SWAP_{12}$ \\
        $CNOT_{21}$ & $SWAP_{13}$ \\
        \hline
        \end{tabular}\\
        \label{tab:ops}
\end{table}
\section{\label{app:projexp}Projection operators for [[4,2,2]]-encoded VQE ansatz}
We use the following projection operators to calculate the logical fidelity of the states corresponding to each post-selection strategy of the [[4,2,2]]-encoded VQE ansatz.
\begin{subequations}
    \begin{align}
        \Pi_A =&|0\rangle\langle0|_{a_1}\otimes|\mathbb{I}_0\mathbb{I}_1\mathbb{I}_2\mathbb{I}_3\rangle\langle\mathbb{I}_0\mathbb{I}_1\mathbb{I}_2\mathbb{I}_3|\otimes |0\rangle\langle0|_{a_2=0}\\
        \Pi_{P} =&\mathbb{I}_{a_1}\otimes(|\overline{00}\rangle\langle\overline{00}|_{q_0-q_3} + |\overline{01}\rangle\langle\overline{01}|_{q_0-q_3} \nonumber\\ &+ |\overline{10}\rangle\langle\overline{10}|_{q_0-q_3} + |\overline{11}\rangle\langle \overline{11}|_{q_0-q_3})\otimes|0\rangle\langle0|_{a_2=0}\\
        \Pi_{AP} =& \Pi_A\otimes \Pi_{P}
    \end{align}
    \label{eq:projector}
\end{subequations}
\section{Projection operators for [[4,2,2]]-encoded input state}
\par
We calculate the fidelity corresponding to each post-selection method for the [[4,2,2]]-encoded input state $|\overline{00}\rangle$ by projecting states (i) with $a_1=|0\rangle$, (ii)  within the codespace, and (iii) both within the codespace and with $a_1=|0\rangle$, using projection operators, $S_A, S_P$, and $S_{AP}$, respectively. We use Equation \ref{eq:fidelity} to calculate the fidelity, $F_{S_i}$, of the projected states where, in this case, $\rho_1 = |\phi\rangle\langle\phi|$ is the noiseless encoded state, $\rho_i = \frac{S_i\rho_{noisy}S_i^{\dagger}}{\Tr(S_i\rho_{noisy}S_i^{\dagger})}$ for $S_i\in\{S_A,S_P,S_{AP}\}$ are the projected states and $\rho_{noisy}$ is the noisy encoded input state. 

\begin{eqnarray}
    |\phi\rangle =&|a_1\rangle \otimes |\overline{00}\rangle\nonumber
    \\=&  |0\rangle \otimes \frac{1}{\sqrt{2}}(|0000\rangle + |1111\rangle)
    \label{eq:sprep}
\end{eqnarray}

\begin{subequations}
    \begin{align}
        S_A &= |0\rangle\langle0|_{a_1}\otimes|\mathbb{I}_0\mathbb{I}_1\mathbb{I}_2\mathbb{I}_3\rangle\langle\mathbb{I}_0\mathbb{I}_1\mathbb{I}_2\mathbb{I}_3|\\
        \begin{split}
        S_P &= \mathbb{I}_{a_1}\otimes(|\overline{00}\rangle\langle\overline{00}|_{q_0-q_3} + |\overline{01}\rangle\langle\overline{01}|_{q_0-q_3}\\ &+ |\overline{10}\rangle\langle\overline{10}|_{q_0-q_3} + |\overline{11}\rangle\langle \overline{11}|_{q_0-q_3})
        \end{split}\\
        S_{AP} &= S_A\otimes S_P
    \end{align}
    \label{eq:spprojector}
\end{subequations}

\bibliography{apstemplate}

\end{document}